\documentclass[12pt,a4paper]{article} 

\usepackage{amsthm}
\usepackage{cancel}
\usepackage{empheq}
\usepackage{tikz}
\usetikzlibrary{shapes.geometric}
\usepackage{float}
\usepackage{blindtext}
\usepackage{enumerate}
\usepackage{enumitem}
\usepackage{amsmath,latexsym,amssymb}
\usepackage{bm}
\usepackage{bbm}
\usepackage{mathtools,slashed}
\usepackage[hidelinks]{hyperref}
\usepackage{graphicx}
\usepackage[labelfont=bf]{caption}
\usepackage{xcolor}
\usepackage[compat=1.1.0]{tikz-feynman}
\usepackage[compat=1.1.0]{tikz-feynhand}
\hypersetup{
 colorlinks,
 linkcolor={blue},
 citecolor={blue},
 urlcolor={blue}
}
\usepackage{appendix}
\numberwithin{equation}{section}

\usepackage[bibstyle=ieee,backend=biber]{biblatex}
\newcommand{\diff}[1]{\mbox{d} #1}
\newcommand{\Tr}[1]{\mbox{Tr} #1}

\newcommand{\sTr}[1]{\mbox{sTr}\, #1}

\def \ione { 
 \begin{tikzpicture}[baseline=-\the\dimexpr\fontdimen22\textfont2\relax]
 \begin{feynhand}
 \vertex [dot] (b) at (0,0) {};
 \vertex[dot](c)at(1.4,0){};
 \draw[line width=.5mm] (b) arc [start angle=180, end angle=-180, radius=.7cm];
 \draw[line width=.5mm] (b) arc [start angle=180, end angle=-180, radius=-.7cm];
 \propag[fer] (b) to [in=90, out=90,looseness=.5] (c);
 \propag[fer] (c) to [in=-90, out=-90,looseness=.5] (b);
 \vertex [ringdot] (d) at (.2,.5) {};
 \vertex [ringdot] (d) at (1.2,.5) {};
 \end{feynhand}
 \end{tikzpicture}
 }

\def \itwo { 
 \begin{tikzpicture}[baseline=-\the\dimexpr\fontdimen22\textfont2\relax]
 \begin{feynhand}
 \vertex [dot] (b) at (0,0) {};
 \vertex[dot](c)at(1.4,0){};
 \draw[line width=.5mm] (b) arc [start angle=180, end angle=-180, radius=.7cm];
 \draw[line width=.5mm] (b) arc [start angle=180, end angle=-180, radius=-.7cm];
 \propag[fer] (b) to [in=90, out=90,looseness=.5] (c);
 \propag[fer] (c) to [in=-90, out=-90,looseness=.5] (b);
 \vertex [ringdot] (d) at (.2,.5) {};
 \vertex [ringdot] (d) at (1.2,-.5) {};
 \end{feynhand}
 \end{tikzpicture}
 }

\def \disc { 
 \begin{tikzpicture}[baseline=-\the\dimexpr\fontdimen22\textfont2\relax]
 \begin{feynhand}
 \vertex [dot] (b) at (0,0) {};
 \vertex [dot] (e) at (1.8,0) {};
 \vertex[dot](c)at(3.2,0){};
 \draw[line width=.5mm] (b) arc [start angle=180, end angle=-180, radius=-.7cm];
 \draw[line width=.5mm] (1.4,0) arc [start angle=-180, end angle=180, radius=-.7cm];
 \propag[fer] (e) to [in=90, out=90,looseness=.5] (c);
 \propag[fer] (c) to [in=-90, out=-90,looseness=.5] (e);
 \propag[fer] (c) to [in=90, out=90,looseness=1.5] (e);
 \propag[fer] (e) to [in=-90, out=-90,looseness=1.5] (c);
 \vertex [ringdot] (d) at (.2,-.5) {};
 \vertex [ringdot] (d) at (-.2,.5) {};
 \end{feynhand}
 \end{tikzpicture}
 }

\def\del{\Delta}
\def\ddel{{}^\bullet\! \Delta}
\def\deld{\Delta^{\hskip -.5mm \bullet}}
\def\dddel{{}^{\bullet \bullet} \! \Delta}
\def\ddeld{{}^{\bullet}\! \Delta^{\hskip -.5mm \bullet}}

\def\lpartial{ \overset{\leftarrow}{\slashed{\partial}}}
\def\rpartial{ \overset{\rightarrow}{\slashed{\partial}}}
\begin{document}

%###############################################################################################
%###############################################################################################
\begin{titlepage} 

\begin{center}
{\LARGE \bf Six-dimensional one-loop divergences in quantum\\[2mm] gravity from the $\mathcal{N}=4$ spinning particle}  
\vskip 1.2cm

Fiorenzo Bastianelli$^{\,a,b}$, Francesco Comberiati$^{\,a,b}$, Filippo Fecit$^{\,a,b}$ and Fabio Ori$^{\,a}$
\vskip 1cm

$^a${\em Dipartimento di Fisica e Astronomia ``Augusto Righi", Universit{\`a} di Bologna,\\
via Irnerio 46, I-40126 Bologna, Italy}\\[2mm]

$^b${\em INFN, Sezione di Bologna, via Irnerio 46, I-40126 Bologna, Italy}\\[2mm]
 
\end{center}
\vskip .8cm

\abstract{In this work, we investigate the computation of the counterterms necessary for the renormalization of the 
one-loop effective action of quantum gravity using both the worldline formalism and the heat kernel method. Our primary contribution is the determination of the Seleey-DeWitt coefficient $a_3(D)$ for perturbative quantum gravity with a cosmological constant, which we evaluate on Einstein manifolds of arbitrary $D$ dimensions. This coefficient characterizes quantum gravity in a gauge-invariant manner due to the on-shell condition of the background on which the graviton propagates. Previously, this coefficient was not fully known in the literature. 
We employ the $\mathcal{N}=4$ spinning particle model recently proposed to describe the graviton in first quantization and then use the heat kernel method to cross-check the correctness of our calculations. Finally, we restrict to six dimensions,
where the coefficient corresponds to the logarithmic divergences of the effective action,
and compare our results with those available in the literature.}

\end{titlepage}
%################################################################################################################
%################################################################################################################

%################################################################################################################
\tableofcontents
%################################################################################################################

%################################################################################################################
%################################################################################################################
\section{Introduction}

One of the most important and challenging areas of modern theoretical physics is the search for a consistent quantum theory of gravity. Regardless of its specific details, any such theory is expected to incorporate and generalize the principles of general relativity. However, as it is well-known, general relativity gives rise to a non-renormalizable quantum field theory, in which diverging terms cannot be absorbed into the parameters of the Einstein-Hilbert action. This makes the development of a quantum theory of gravity an ongoing and active area of research, with important implications for our understanding of the fundamental nature of spacetime.

The investigation of divergences in quantum gravity dates back to the pioneering work of t' Hooft and Veltman \cite{tHooft:1974toh}. They demonstrated that, when evaluated on-shell and with vanishing cosmological constant, the one-loop logarithmic divergences in four dimensions were absent. Subsequently, van Nieuwenhuizen \cite{VanNieuwenhuizen:1977ca} observed that one-loop quantum calculations in six dimensions share some features with two-loop calculations in four dimensions. He found that quantum gravity in six dimensions contains a nonvanishing logarithmic divergence, suggesting that divergences could emerge at two-loops in four dimensions as well. This prediction was supported by Critchley \cite{Critchley:1978kb}, who corrected a numerical factor in the six-dimensional term. Eventually, Goroff and Sagnotti \cite{Goroff:1985th} explicitly calculated the two-loop divergence in four dimensions. They demonstrated that pure quantum gravity is a non-renormalizable theory at two-loops. This result was later checked and confirmed by van de Ven \cite{vandeVen:1991gw}. On the other hand, the inclusion of a cosmological constant gives rise to a one-loop logarithmic divergence already in four dimensions, as found by Christensen and Duff \cite{Christensen:1979iy}.

Subsequent research has focused on the search and analysis of gravitational theories with improved ultraviolet behaviour, such as simple supergravity \cite{Freedman:1976xh,Deser:1976eh} and supergravities with extended supersymmetry, like the $N= 8$ supergravity \cite{Cremmer:1979up} that was obtained by dimensional reduction of the unique 11D supergravity \cite{Cremmer:1978km}. The ultraviolet properties of these theories are still under investigation to understand whether all supergravities must necessarily be ultraviolet divergent, as known symmetry arguments seem to suggest, see \cite{Bern:2018jmv} and references therein. On the other hand, the various supergravity theories appear as low energy limits of string theory \cite{Green:2012oqa, Polchinski:1998rq}, whose finiteness is related to its being a theory of extended objects rather than point-like particles. 

In this work, we focus on pure gravity with cosmological constant with the aim to investigate further the structure of the diverging terms in the one-loop effective action. Evaluated on-shell, these diverging terms are gauge-invariant and characterize unambiguously the theory. For instance, they could serve as a benchmark for verifying alternative approaches to perturbative quantum gravity, and thus their precise expression should be known explicitly. Our main contribution is the determination of the Seleey-DeWitt coefficient $a_3(D)$ of perturbative quantum gravity that parametrizes a class of divergences that start to appear in $D\geq 6$ dimensions. It has not been reported in the literature in its full generality so far. On the other hand, the coefficients $a_n(D)$ for $n=0,1,2$ are already known and have been cross-checked with different methods \cite{Bastianelli:2019xhi, Bastianelli:2022pqq, Bastianelli:2023oyz}.

In our endeavor to determine the coefficient $a_3(D)$, we employ two distinct approaches: the worldline formalism and the heat kernel method. The first one consists in using the $\mathcal{N} = 4$ spinning particle, which provides a first-quantized description of the graviton. It correctly describes the graviton propagating on Einstein manifolds \cite{Bonezzi:2018box} and it was employed in \cite{Bastianelli:2019xhi} to construct a worldline representation of the one-loop effective action for quantum gravity. The worldline representation involves computing the path integral of the $\mathcal{N} = 4$ spinning particle on a circle. We perform the perturbative expansion of this path integral and compute it up to the order required to determine the coefficient $a_3(D)$. Then, we employ a second method based on the time-honored heat kernel representation of the one-loop effective action of quantum gravity \cite{DeWitt:1964mxt, DeWitt:1984sjp, DeWitt:2003pm}. In this approach, we evaluate the heat kernel coefficients by taking on-shell the background metric, i.e. inserting metrics corresponding to Einstein spaces. This second method yields the same coefficient $a_3(D)$ obtained previously. The agreement between the two methods provides a robust consistency check for both the new coefficient and the worldline $\mathcal{N} = 4$ representation.

The paper is organized as follows. In section \ref{sec2}, we provide a description of the worldline formalism. Specifically, we focus on the path integral for the $\mathcal{N} = 4$ spinning particle and show how it can be used to compute the on-shell counterterms in quantum gravity. We compute the explicit counterterms related to the coefficients $a_n(D)$ for $n=0,1,2,3$ using worldline perturbation theory. Section \ref{sec3} employs the heat kernel method in one-loop quantum gravity to derive the same counterterms, thus providing a cross-check of our results. In section \ref{sec4}, we discuss the one-loop divergences for quantum gravity in four and six dimensions, providing further consistency with the existing literature. Finally, our conclusions are presented in section \ref{sec5}. The appendices contain useful formulae for geometric quantities on Einstein spaces and detailed explanations of the computational procedures employed.
 
%################################################################################################################ 
%################################################################################################################
\section{Worldline formalism} \label{sec2}

In this section, we review the representation of the one-loop effective action for pure Einstein-Hilbert gravity in the worldline formalism. The most elegant way of obtaining such a representation is to consider a relativistic spinning particle with four local supersymmetries on the worldline, check that its spectrum coincides with that of the graviton in $D$ dimensions --- especially when the coupling to a background metric is introduced --- and then path integrate the model on the circle to obtain the desired one-loop effective action for the graviton as a functional of the background metric. Along the way, one finds that quantum consistency of the model requires the background metric to satisfy Einstein's equations of motion. This approach for describing the graviton shares many analogies with string theory: they are both first quantized models and both of them require the allowed background fields to be on-shell. We shall not present here all the details leading to the construction of the worldline representation of the effective action for pure quantum gravity, for which we refer to \cite{Bastianelli:2019xhi}, but we review the main ideas that have led to that result.
 
Massless relativistic particles carrying spin $s$ have a description in terms of mechanical models with ${\cal N} = 2s$ local supersymmetries on the worldline, as suggested in \cite{Berezin:1976eg} and explicitly constructed in \cite{ Gershun:1979fb, Howe:1988ft}. These models are described by the worldline particle coordinates $x^\mu$ together with ${\cal N}$ real fermionic superpartners $\psi^\mu_i$ with $i=1,\dots,{\cal N}$, 
introduced to describe the spin degrees of freedom of the particle. Unitarity of the model requires the $\mu=0$ components of these variables to be non-physical. The ${\cal N}$-extended local supersymmetry is
needed precisely to compensate for this redundancy in a relativistic invariant manner. A further gauging of the R-symmetry group SO($\cal N$) that rotates the $\cal N$ supercharges is optional as it is not required by unitarity. It is generically used to constrain the model to have the minimal amount of degrees of freedom and deliver pure spin $s$ states. This model was path integrated on a circle in \cite{Bastianelli:2007pv,} to verify that it propagates correctly the degrees of freedom of a relativistic particle of spin $s$. While everything works fine in flat spacetimes, coupling to background fields, and in particular to curved spacetimes, proved more difficult to achieve, which is somehow expected for particles of sufficiently high spin. Early results were obtained in \cite{Kuzenko:1995mg, Bastianelli:2008nm,}, where some obstructions were circumvented allowing for specific couplings to (A)dS and conformally flat spaces. Then, a crucial result was obtained in \cite{Dai:2008bh}, where it was shown how to use BRST methods to extend the case of spin 1 to include non-abelian couplings, thus providing a first-quantized description of the gluon. Using similar BRST techniques, in \cite{Bonezzi:2018box} it was found how the spin 2 massless particle could be coupled to background metrics that satisfy Einstein's equation. This guarantees that the first-quantized graviton can propagate consistently on Einstein spacetimes, where the Ricci tensor is proportional to the metric 
\begin{equation} \label{1EM}
R_{\mu\nu} = \lambda g_{\mu\nu}
\end{equation}
with constant $\lambda$, thus admitting a cosmological constant of indefinite sign. The BRST construction was crucial to develop a path integral quantization delivering a worldline representation of the one-loop effective action for quantum gravity \cite{Bastianelli:2019xhi}. Path integrals for one-dimensional nonlinear sigma models, such as the one associated with the $\mathcal{N} = 4$ spinning particle in curved spaces, require counterterms that are related to the regularization scheme used in defining the path integral itself \cite{Bastianelli:2006rx, new-book}. The counterterm used in \cite{Bastianelli:2019xhi} was the one associated with wordline dimensional regularization and was effectively valid only for $D=4$. Its extension to arbitrary $D$ dimensions was constructed in \cite{Bastianelli:2022pqq} and it is the one that we use here below. 

Alternative worldline representations of the effective action for quantum gravity are possible, see for instance \cite{Bastianelli:2013tsa, Bastianelli:2023oyz}. They are close in spirit to the heat kernel approach
employed in Section \ref{sec4}. For their formulation they need direct inputs from the associated QFT, just as it happens in the heat kernel method. In this sense, they are not independent of the second quantized theory 
and will not be used in this section. 

As anticipated, we refer to the aforementioned studies for details on the analysis of the spinning particle action and quantization, proceeding now with the construction of the worldline representation of the gravitational effective action.
 
%################################################################################################################
\subsection{The worldloop path integral}
%################################################################################################################

The one-loop effective action $\Gamma [g_{\mu\nu}]$ for pure gravity corresponds to the path integral of the $\mathcal{N}=4$ spinning particle action $S[X,G;g_{\mu\nu}]$ on worldlines with the topology of the circle $S^1$, also called the ``worldloop'', and takes the schematic form 
\begin{equation} \label{1}
\Gamma[g_{\mu\nu}] = \int_{S^1}
\frac{\mathcal{D}G\,\mathcal{D}X}{\mathrm{Vol(Gauge)}}\, {\rm e}^{-S[X,G;g_{\mu\nu}]}\ .
\end{equation}
The particle Euclidean action depends on the worldline gauge fields $G=\left( e, \chi, \bar{\chi},a\right)$ and coordinates with supersymmetric partners $X=\left(x,\psi,\bar{\psi}\right)$, while the overcounting from summing over gauge equivalent configurations is formally taken into account by dividing by the volume of the gauge group. The four real fermionic partners of the coordinates have here been cast more conveniently into a pair of complex fermionic variables $\psi$ and $\bar \psi$, as well as the worldline gravitinos $\chi$ and $ \bar \chi$. As implied by the BRST analysis, the gauge symmetries are consistent only when the background 
metric $g_{\mu\nu}$ is on-shell, i.e. satisfies eq. \eqref{1EM}. Explicitly, the effective action \eqref{1} is related to a partially gauge-fixed version of the $\mathcal{N}=4$ spinning particle path integral $Z(T)$ through a Schwinger representation, and given by 
\begin{equation} \label{euc}
\Gamma[g_{\mu\nu}] = -\frac{1}{2}\int_{0}^{\infty}\frac{dT}{T}Z(T)
\end{equation}
where $T$ is often called the Schwinger proper time, a modulus whose integration arises from the gauge-fixing of the einbein $e$ on the circle. In the present work, we only mention some of the important technicalities, namely the gauging of a parabolic subgroup of the $SO(\mathcal{N})$ $R$-symmetry group, the choice of the aforementioned gauge fixing of the worldline action, and the regularization of the nonlinear supersymmetric sigma model, skipping the details while referring to previous work for the reader interested in them \cite{Bastianelli:2019xhi, Bastianelli:2022pqq}. We choose to focus our discussion on the perturbative computation of the path integral providing only a few remarks when needed. The partition function $Z(T)$, upon gauge fixing, takes the following explicit form 
\begin{align} \label{master}
Z(T)&=\int_{0}^{2\pi}\frac{d\theta}{2\pi}\int_{0}^{2\pi}\frac{d\phi}{2\pi}\;P(\theta,\phi) \;\int_{_{\rm PBC}}\hskip-.4cm{ D}xDaDbDc\int_{_{\rm ABC}}\hskip-.4cm D\bar{\psi}D\psi \, e^{-S[X;g_{\mu\nu}]}\ ,
\end{align}
where $P(\theta,\phi)$ is the measure on the moduli space $(\theta, \phi)$ generated by the gauge fixing and which implements the correct projection on the physical graviton Hilbert space in $D$ dimensions. The worldline variables $X=\left(x,\psi,\bar{\psi},a,b,c\right)$ now include bosonic $a$ and fermionic $(b,c)$ ``metric ghosts'', introduced in order to keep translational invariance of the path integral measure and which renormalize potentially divergent worldline diagrams \cite{Bastianelli:2006rx}. The path integral over bosonic variables and metric ghosts is evaluated by fixing periodic boundary conditions (PBC), while the fermionic path integral is performed by choosing antiperiodic boundary conditions (ABC) on each flavor of fermionic fields $\psi^a_{i}$, with the internal index $i$ taking values $i=1,2$.\footnote{Due to the complex combination of the original $\mathcal{N}=4$ real fermions.} The gauge-fixed nonlinear sigma model action reads\footnote{The bosonic coordinates are understood to be shifted as $\dot{x}^{\mu}\dot{x}^{\nu}\rightarrow\dot{x}^{\mu}\dot{x}^{\nu}+ a^{\mu}a^{\nu}+b^{\mu}c^{\nu}$. This shift implements into \eqref{action} the ghost action $S_{\rm{gh}}[x,a,b,c]=\int d\tau \tfrac{1}{4T} g_{\mu\nu}(x)(a^{\mu}a^{\nu}+b^{\mu}c^{\nu})$ which allows for the exponentiation of the determinant factor hidden inside the path integral measure on a curved spacetime, i.e. 
\begin{equation*}
\mathcal{D}x=\prod_\tau d^Dx(\tau)\sqrt{g(x(\tau))}={D}x\int Da Db Dc
\; e^{-S_{\rm{gh}}}\ ,
\end{equation*}
where $Dx$, $Da$, $Db$, and $Dc$ are the standard translational invariant measures. In particular, these metric ghosts create worldline divergences that compensate for the divergences generated by correlators of the $\dot x^\mu$'s. Divergences formally cancel out and one is left with a finite theory, whose remaining ambiguities are taken care of by choosing a regularization scheme with a corresponding counterterm that remains finite.}
\begin{align} \label{action}
S[X;g_{\mu\nu}]=\int d\tau \Big[ &\frac{1}{4T}g_{\mu\nu}(x)\,\dot{x}^{\mu}\dot{x}^{\nu}+\bar{\psi}^{a i}\left(
	\delta_i^j D_\tau - \hat a_{i}^{j}\right){\psi}_{aj} -T R_{abcd}(x) \, \bar{\psi}^{a}\cdot \psi^{b} \bar{\psi}^{c}\cdot \psi^{d} -T \, \mathcal{V}(x)\Big]\ ,
\end{align}
where we use flat indices on the worldline complex fermions $\psi^{a}_{i}$ and denoted the covariant derivative with spin connection $\omega_{\mu ab}$ acting on the fermions by $D_\tau \psi^a_i = \partial_\tau \psi^a_i + \dot x^\mu \omega_{\mu}{}^a{}_b (x) \psi^b_i$. We also used a dot to indicate contraction on the internal indices and denoted
\begin{equation}
\hat a_{i}^{j}=\left(\begin{array}{cc} \theta & 0 \\ 0 & \phi \\
\end{array}\right)
\end{equation}
the gauge-fixed values of the worldline gauge fields acting on the fermions and related to the gauging of the parabolic subgroup of the $R$-symmetry group. The angles $\theta$ and $\phi$ are precisely the two leftover moduli remaining after the gauge-fixing procedure.\footnote{As discussed in \cite{Bastianelli:2019xhi}, the gauging of the parabolic subgroup allows the one-loop measure for the path integral to be modified so that it projects \emph{exactly} onto the graviton state. Alternatively, the gauging of the entire SO($4$) group would result in the graviton plus unwanted contributions of topological nature, a case that might be worth studying but is beyond the scope of our work.}

A few comments are in order. The theory described by \eqref{action} should be seen as a one-dimensional field theory living on the worldline with the bosonic fields $x^\mu(\tau)$, the embedding of the worldline into spacetime, taking values in a $D$-dimensional target space $\cal{M}$. On the worldline, one usually finds it convenient to rescale the parameter $\tau$ to take values on the finite interval $I \equiv [0, 1]$. A crucial role is played by the scalar potential term $\mathcal{V}$ of quantum origin \cite{Bastianelli:2022pqq}. It is necessary since it contains the counterterm required by the regularization scheme one decides to use to define the path integral and an additional potential needed to achieve nilpotency of the BRST charge at the quantum level. The latter condition requires a value of $V_{\rm BRST}= \frac{2}{D} R$ in the Hamiltonian constraint 
 \cite{Bonezzi:2018box}. Regarding the former, in the present work we adopt dimensional regularization (DR) on the worldline, as used for instance in \cite{Bastianelli:2000dw, Bastianelli:2002fv, Bastianelli:2002qw}
 in similar contexts, while reading from \cite{Bastianelli:2011cc} the counterterm $V_{\rm CT}=-\frac{1}{4} R$ needed for the case of four supersymmetries, thus producing an effective potential
\begin{equation}
\mathcal{V}=V_{\rm BRST}+V_{\rm CT}=\left(\frac{2}{D}-\frac{1}{4}\right)R \equiv \Omega R
\end{equation}
which indeed is the one used in \cite{Bastianelli:2022pqq}.
Finally, for computational purposes, it is convenient to rewrite the angular integrations over the moduli $\theta$ and $\phi$ in the complex plane. This can be achieved by introducing the Wilson variables $z\equiv e^{i\theta}$ and $\omega\equiv e^{i\phi}$, so to recast the partition function as
\begin{align} \label{MS} 
Z(T)&=\oint \frac{dz}{2\pi i}\frac{d\omega}{2\pi i}\;P(z,\omega) \;\int_{_{\rm PBC}}\hskip-.4cm{ D}xDaDbDc\int_{_{\rm ABC}}\hskip-.4cm D\bar{\psi}D\psi \, e^{-S[X;g_{\mu\nu}]}\ ,
\end{align}
where the modular integration is performed over the circle $|z|=1$, with the singular point $z=-1$ pushed out of the contour,\footnote{Poles at $z=-1$ arise when computing perturbative corrections and are excluded by this prescription. The same goes for $\omega$. Discussion on the regulated contour of integration for the modular parameters can be found in \cite{Bastianelli:2005vk}.} and with the measure on the moduli space being now
\begin{equation} \label{measure}
P(z,\omega)=\frac{1}{2}\frac{(z+1)^{D-2}}{z^{3}}\frac{(\omega+1)^{D-2}}{\omega^{3}}(z-\omega)^{2}(z\omega-1)\ .
\end{equation}
In the next section, we set up the perturbative expansion for small values of $T$ of the path integral \eqref{MS}, such that the full $T^{n}$ correction is an $(n+1)$-loop expansion in the worldline theory, which will allow us to identify the divergences in the effective action of pure gravity.

%################################################################################################################
\subsection{Setting up the perturbative expansion}
%################################################################################################################

Having at hand a path integral representation for the effective action, it is possible to set up the perturbative expansion around the free theory. However, there are two issues to take care of in order to be able to perform calculations. The first involves factorizing out the zero modes in the kinetic operator in \eqref{action}. Zero modes appear perturbatively once expanding around a constant metric and when considering periodic boundary conditions. To this task, we parametrize the bosonic coordinates of the circle (interpreted as a parametrization of the particle paths in target space) as
\begin{equation}
x^{\mu}(\tau) = x_{0}^{\mu}+q^{\mu}(\tau)\ ,
\end{equation}
thus describing all loops in spacetime with a fixed base point $x^{\mu}_{0}$ (the zero mode that is integrated over only at the end) plus quantum fluctuations with vanishing Dirichlet boundary conditions (DBC), 
indicated by $q^{\mu}(\tau)$ and thus satisfying $ q^\mu(0) =q^\mu(1) =0$. Note that the fermionic coordinates have no zero modes, due to their antiperiodic boundary conditions. The second issue consists in expanding in Riemann normal coordinates (RNC) centered around $x^{\mu}_{0}$, so to write the metric tensor and the spin connection as follows \cite{Bastianelli:2000dw, Muller:1997zk}
\begin{align} 
\begin{split}
g_{\mu\nu}(x(\tau)) &= g_{\mu\nu} + \frac13 R_{\alpha\mu\nu\beta} q^\alpha q^\beta +\frac16 \nabla_\gamma R_{\alpha\mu\nu\beta} q^\alpha q^\beta q^\gamma+ R_{\alpha\beta\mu\nu\gamma\delta} q^\alpha q^\beta q^\gamma q^\delta \\
&\phantom{=}+\frac{1}{315}R_{\mu\alpha\beta}{}^{\sigma}R_{\sigma\gamma\delta}{}^{\lambda}R_{\lambda\tau\epsilon\nu}q^\alpha q^\beta q^\gamma q^\delta q^\tau q^\epsilon +\mathcal{O}(q^6) \label{RNCg} 
\end{split}\\[.5em]
\begin{split}
\omega_{\mu ab}(x(\tau)) &= \frac12 R_{\alpha\mu ab} q^\alpha +\frac13 \nabla_\alpha R_{\beta\mu ab} q^\alpha q^\beta +\frac18 \nabla_\alpha \nabla_\beta R_{\gamma\mu ab} +\frac{1}{24}R^{\tau}{}_{\alpha\beta\mu} R_{\gamma\tau ab}q^\alpha q^\beta q^\gamma + \mathcal{O}(q^4) \label{RNCw}
\end{split}
\end{align}
where \begin{equation}
R_{\alpha\beta\mu\nu\gamma\delta} = \frac{1}{20} \nabla_\delta \nabla_\gamma R_{\alpha\mu\nu\beta} +\frac{2}{45}R_{\alpha\mu}{}^\sigma{}_\beta R_{\gamma\sigma\nu\delta}
\end{equation}
and where we only kept the terms needed to obtain a perturbative expansion to order $T^3$. In \eqref{RNCg}-\eqref{RNCw} and henceforth, unless specified otherwise, we intend all tensor structures to be evaluated at the initial point $x^{\mu}_{0}$, thus factorizing out their dependence upon the worldline bosonic variables $q^\mu(\tau)$. Finally, the Riemann tensor appearing in the four-fermions interaction in \eqref{action} has to be Taylor expanded as well, around the same reference point $x^{\mu}_{0}$. The perturbative expansion of the path integral \eqref{MS} reads 
\begin{equation} \label{4}
Z(T)=\oint \frac{dz}{2\pi i}\frac{d\omega}{2\pi i}\;P(z,\omega) \;\int d^{D}x_{0}\frac{\sqrt{g(x_{0})}}{\left(4\pi T\right)^{\frac{D}{2}}}\Big\langle e^{-S_{\rm int}}\Big\rangle \ ,
\end{equation}
factorizing out for convenience the $\sqrt{g(x_{0})}$ arising from the free ghost part of the action after functional integrating, together with $\left(4\pi T\right)^{-\frac{D}{2}}$ arising from the free particle 
path integral. The remaining expectation value is to be evaluated using the Wick theorem on the free path integral, with the free action being given by the quadratic part of \eqref{action}, namely
\begin{equation} \label{freeaction}
S_{0}[X] = \int d\tau \left[\frac{1}{4T}g_{\mu\nu}\left( \dot{x}^{\mu}\dot{x}^{\nu}+ a^{\mu}a^{\nu}+b^{\mu}c^{\nu}	\right)+\bar{\psi}^{ai}\left(\delta_i^j \, \partial_{\tau}-\hat a_{i}\,^{j}\right)\psi_{aj}\right]\ ,
\end{equation} 
from which one obtains the worldline propagators of the theory, reported in appendix \ref{appendixB1}. Higher order terms form the interacting action $S_{\rm int}$, to be analyzed later on. It is possible to recast \eqref{4} in a more compact form introducing the double expectation value of the interacting action $\langle \hskip -.05 cm \langle\cdots \rangle \hskip -.05cm \rangle$, namely the average over the path integral and over the moduli space parametrized by $z$ and $w$
\begin{equation}
\Big\langle \hskip -.1cm \Big\langle e^{-S_{\rm int}}\Big\rangle \hskip -.1cm \Big\rangle
 = \oint \frac{dz}{2\pi i}\frac{d\omega}{2\pi i}\;P(z,\omega) \Big\langle e^{-S_{\rm int}}\Big\rangle \label{double}\ .
\end{equation}
Identifying the expectation values defined above as
\begin{align} 
\Big\langle e^{-S_{\rm int}}\Big\rangle = \sum_{n=0}^{\infty}a_{n}(D,z,\omega)\, T^{n} \quad \longrightarrow \quad \Big\langle \hskip -.1cm \Big\langle e^{-S_{\rm int}}\Big\rangle \hskip -.1cm \Big\rangle 
= \sum_{n=0}^{\infty}a_{n}(D)\, T^{n}\ , \label{2.16}
\end{align}
allows us to rearrange the path integral \eqref{4} so as to make explicit the Seeley-DeWitt coefficients arising from the perturbative expansion
\begin{align}
Z(T)=\int \frac{d^{D}x_{0}}{\left(4\pi T\right)^{\frac{D}{2}}}\sqrt{g(x_{0})}\left[a_0(D)
+ a_1(D) \,T + a_2 (D)\, T^2 + a_3 (D)\, T^3+ \mathcal{O}(T^{4})\right]\ .\label{series}
\end{align}
One can recognize that, while in the above sum $a_{0}(D,z,\omega)=1$, its projected partner 
\begin{equation}
a_{0}(D) = \langle \hskip -.05cm \langle 1 \rangle \hskip -.05cm \rangle =
\frac{1}{2}\oint \frac{dz}{2\pi i}\frac{d\omega}{2\pi i}\frac{(z+1)^{D-2}}{z^{3}}\frac{(\omega+1)^{D-2}}{\omega^{3}}(z-\omega)^{2}(z\omega-1) = \frac{D(D-3)}{2} 
\end{equation}
gives the massless graviton physical polarizations in $D$ spacetime dimensions when using the correct measure $P(z,\omega)$ in eq. \eqref{measure}. We have now made explicit our main task, namely to determine the Seeley-DeWitt coefficient $a_3(D)$ in the perturbative expansion on Einstein spaces.

%################################################################################################################
\subsection{Outline of the computation} \label{sec2.3}
%################################################################################################################

We will now give a brief outline of the procedure, delegating the details of both the computation and regularization of potentially divergent diagrams to appendix \ref{appendixB}. To systematically work out all perturbative contributions to the desired order, one has first to identify and compute the connected worldline diagrams arising from the path integral expansion. In order to do so, we report the interacting action in \eqref{4} expanded to the desired order
\begin{align} \label{S-int}
\begin{split}
S_{\rm int}&=\int d\tau\Bigg[
 \frac{1}{4T}\bigg(\frac{1}{3}R_{\alpha\mu\nu\beta}q^\alpha q^\beta +\frac{1}{6}\nabla_\gamma R_{\alpha\mu\nu\beta}q^\alpha q^\beta q^\gamma+\frac{1}{20} \nabla_\delta\nabla_\gamma R_{\alpha\mu\nu\beta}q^\alpha q^\beta q^\gamma q^\delta\\
&+\frac{2}{45}R_{\alpha\mu}{}^{\sigma}\,_{\beta}R_{\gamma\sigma\nu\delta}q^\alpha q^\beta q^\gamma q^\delta+\frac{1}{315}R_{\mu\alpha\beta}{}^{\sigma}R_{\sigma\gamma\delta}{}^{\lambda}R_{\lambda\tau\epsilon\nu}q^\alpha q^\beta q^\gamma q^\delta q^\tau q^\epsilon \bigg)\left(\dot{q}^\mu\dot{q}^\nu +a^\mu a^\nu +b^\mu b^\nu \right)\\
&+ 
\left( \frac{1}{2}R_{\alpha\mu ab}q^\alpha+\frac{1}{3}\nabla_\alpha R_{\beta\mu a b}q^\alpha q^\beta +\frac{1}{8}\nabla_\alpha \nabla_\beta R_{\gamma\mu a b}q^\alpha q^\beta q^\gamma+\frac{1}{24}R^{\tau}{}_{\alpha\beta\mu}R_{\gamma\tau a b }q^\alpha q^\beta q^\gamma\right)
\bar{\psi}^a \cdot \psi^b\, \dot{q}^\mu 
\\
&-T\left(R_{abcd}+q^\alpha \nabla_\alpha R_{abcd}+\frac{1}{2}q^\alpha q^\beta \nabla_\alpha \nabla_\beta R_{abcd} \right)\bar{\psi}^{a}\cdot \psi^{b}\bar{\psi}^{c}\cdot \psi^{d}-T\,\Omega R\Bigg]\ ,
\end{split}
\end{align}
which we write in a more compact form as 
\begin{align}
\begin{split} 
S_{\rm int}=&\frac{1}{4T}\left(S_{\rm K1} +DS_{\rm K1} + D^2S_{\rm K1} +S_{\rm K2} +S_{\rm K3}\right)\\ &
+ S_{\rm C1}+ DS_{\rm C1} + D^2S_{\rm C1}+ S_{\rm C2}\\
& -T\left( S_{\rm F}+ DS_{\rm F} +D^2 S_{\rm F}\right)+TS_{\rm V}\ , \label{10}
\end{split}
\end{align}
where the explicit expression of each term is obtained by comparing with the previous expression, see also appendix \ref{appendixB}. Expanding the exponential in the path integral to order $T^3$, we obtain the contributions that need to be path-averaged and computed using Wick contractions. We list them here, leaving their systematic analysis to appendix \ref{appendixB2}, where we also show more details on the intermediate steps of computation:
\begin{align}
\begin{split} \label{terms}
e^{-S_{\rm int}} \Big|_{T^{3}}&=S_{\rm KIN}-\frac{1}{8T}S_{\rm K1}S_{\rm C1}^{2}+S_{\rm C1}S_{\rm C2}+\frac{1}{2}DS_{\rm C1}^{2}+D^{2}S_{\rm C1}S_{\rm C1}\\
&\phantom{=}+\frac{1}{2}T^{2}S_{\rm C1}^{2}S_{\rm F}+\frac{1}{6}T^{3}S_{\rm F}^{3}+\frac{1}{2}T^2DS_{\rm F}^2 +T^2D^{2}S_{\rm F}S_{\rm F}\ ,
\end{split}
\end{align}
where we have collectively denoted $S_{\rm KIN}$ the three-loop contributions arising from the pure kinetic term, see \eqref{B27}. As an illustrative example, we shall show how to compute a contribution containing all the main features of this type of calculation, namely the correction arising from the spin-connection vertex only\footnote{The subscript $01$ in \eqref{C1C2} (as well as in appendix \ref{appendixB}) serves as a shorthand notation to indicate henceforth $\int_{01}\equiv \int_{0}^{1} d\tau\int_{0}^{1} d\sigma$ in worldline integrals.}
\begin{align} \label{C1C2}
\langle S_{\rm C1} S_{\rm C2}\rangle =\frac{1}{48}R_{\alpha\mu ab}R^{\tau}{}_{\beta\lambda\nu}
R_{\rho \tau cd}\int_{01}\langle \dot{q}^{\mu}_{0}q^{\alpha}_{0}\dot{q}^{\nu}_{1}q^{\beta}_{1}q^{\lambda}_{1}q^{\rho}_{1}\rangle \langle \bar{\psi}^{a}_{0}\cdot \psi_{0}^{b} \bar{\psi}^{c}_{1}\cdot \psi_{1}^{d}\rangle\ .
\end{align}
The fermionic Wick contractions produce
\begin{equation} 
R_{\alpha\mu ab}R^{\tau}{}_{\beta\lambda\nu}
R_{\rho \tau cd} \, \langle \bar{\psi}^{a}_{0}\cdot \psi_{0}^{b} \bar{\psi}^{c}_{1}\cdot \psi_{1}^{d}\rangle
=-\del_{\rm{AF}}{}_{ji}(\tau,\sigma)\del_{\rm{AF}}{}_{ij}(\sigma,\tau)R_{\alpha\mu ab}R^{\tau}{}_{\beta\lambda\nu}R_{\rho\tau}{}^{ba}\ ,
\end{equation}
where we introduced the $\mathcal{N}=4$ fermionic propagator $\del_{\rm{AF} \,}$. Evaluating then the bosonic contractions one gets different Riemann tensor strings, which can be reduced using Bianchi identity, namely
\begin{align}
R^{\mu\nu\rho\sigma}R_{\mu\nu}{}^{\alpha\beta}R_{\rho\alpha\sigma\beta}=\frac{1}{2}R^{\mu\nu\rho\sigma}R_{\mu\nu}{}^{\alpha\beta}R_{\rho\sigma\alpha\beta}\;.
\end{align}
Recall that all the calculations have to be further carried out with two precautions, namely to use Einstein spaces simplifications (see appendix \ref{appendixA}) and to perform the regularization using DR. Then, collecting all terms one gets 
\begin{equation}
\left( \frac{1}{4}R_{\mu\nu\rho\sigma}R^{\rho\sigma\alpha\beta}R_{\alpha\beta}{}^{\mu\nu}+\frac{R}{6 D} R_{\mu\nu\rho\sigma}^{2}\right) \left(
\ione -\itwo +z\rightarrow \omega 
\right)
\end{equation}
where in the graphical representation of the worldline Feynman diagrams full dots denote vertices, an empty dot represents a derivative, a line denotes a bosonic propagator, and oriented lines represent $z$-fermionic propagators. The first diagram has to be regularized due to the singularity carried by the bosonic propagator $\ddeld$. In DR it has to be $d$-dimensional extended as outlined in the following:
\begin{align}
& \hspace{-1cm}
\ione = \int d\tau d\sigma \ \ddeld (\tau,\sigma) \del(\tau,\sigma) \del(\tau,\tau) 
F(z,\tau,\sigma)F(z,\sigma,\tau) 
\nonumber \\
& \hspace{-.9cm}
\rightarrow \int d^{d+1}t\, d^{d+1}s \, 
{}_{\alpha}\del_{\beta}(t,s)\, \del(s,t)\del(t,t)\Tr\left( \gamma^{\alpha}F(z,t,s)\gamma^{\beta}F(z,s,t)\right)
\nonumber \\
=-&\int d^{d+1}t\, d^{d+1}s \ 
{}_{\alpha}\del(t,s)\, \del(t,s)_{\beta}\del(t,t)\Tr\left(\gamma^{\alpha} F(z,t,s)\gamma^{\beta}F(z,s,t)\right) 
\nonumber \\
-&\int d^{d+1}t\, d^{d+1}s \
{}_{\alpha}\del(t,s)\del(s,t)\del(t,t)\Tr \left(F(z,t,s)\lpartial F(z,s,t)+F(z,t,s)\rpartial F(z,s,t) \right)
\nonumber \\
\rightarrow -&\int d\tau d\sigma \, \ddel(\tau,\sigma)\deld(\tau,\sigma)\del(\tau,\tau) F(z,\tau,\sigma)F(z,\sigma,\tau)
\nonumber \\
 \hskip.3cm = -\hskip .1cm &\itwo =\frac{z}{60(z+1)^{2}} \label{2.32}
\end{align}
where in reaching the last-but-one line we used the regulated Green equation for the component $F$ of the fermionic propagator $\del_{\rm{AF} \,}$, see eqs. \eqref{AF}, \eqref{B16} and \eqref{greenEQ2} of appendix \ref{appendixB}, and then removed the regularization by sending the extra dimensions to zero ($d\to 0$). Finally, taking into account the $\omega$ partner of the integrals in \eqref{2.32}, one gets the following result:
\begin{align}
\langle S_{\rm C1}S_{\rm C2}\rangle = \left( \frac{1}{120}R_{\mu\nu\rho\sigma}R^{\rho\sigma\alpha\beta}R_{\alpha\beta}{}^{\mu\nu}+\frac{R}{180D} R_{\mu\nu\rho\sigma}^{2}\right)\left(\frac{z}{(z+1)^{2}}+\frac{\omega}{(\omega+1)^{2}}\right)\;.
\end{align}
Once having worked out systematically all the corrections up to and including order $T^3$ related to the connected graphs, it only remains to exponentiate them and, finally, Taylor expand in $T$ so as to reach the desired order and get the full result.

%################################################################################################################
\subsection{Seeley-DeWitt coefficients} \label{sec2.4}
%################################################################################################################

The calculation of the various terms in the perturbative expansion delivers the following coefficients in the perturbative series \eqref{series}, including the newly found $a_3(D)$
\begin{align}
\begin{split}
a_{0}(D)&= \frac{D(D -3)}{2} \label{a0}
\end{split}\\[.5em]
\begin{split}
\label{a1} a_{1}(D)&= \frac{D^{2}-3D-36}{12} \; R
\end{split}\\[.5em]
\begin{split}
\label{a2} a_{2}(D)&=
\frac{5 D^3 -17 D^2 -354 D -720}{720D }\; R^2 +\frac{D^2-33 D+540 }{360} \; R_{\mu\nu\rho\sigma}^{2}
\end{split}\\[.5em]
\begin{split}
a_{3}(D)&= \frac{35 D^4-147 D^3-3670 D^2-13560D-30240}{90720 D^2}\;R^3+\frac{7 D^3-230 D^2+3357 D+12600}{15120 D}\;R\,R_{\mu\nu\rho\sigma}^{2}\\[.3em]
&\phantom{=}+ 
\frac{17 D^2-555 D-15120}{90720} \;R_{\mu\nu\rho\sigma}R^{\rho\sigma\alpha\beta}R_{\alpha\beta}{}^{\mu\nu}+\frac{D^2-39 D-1080}{3240}\; R_{\alpha\mu\nu\beta}R^{\mu\rho\sigma\nu}R_{\rho}{}^{\alpha\beta}{}_{\sigma}\ . \label{a3}
\end{split}
\end{align}
The above expressions are understood to be gauge-invariant, as they have been calculated specifically on Einstein spaces for the reasons previously discussed. As we shall discuss later, the newly computed coefficient $a_3(D)$ plays a key role in quantum gravity in six dimensions, as it is related to the logarithmic divergences, hence it will be useful to read its value explicitly in this case as well. In $D=6$ it reduces to
\begin{align}
\left. a_{3}\right|_{D=6}&=-\frac{799}{11340}\; R^3 +\frac{481}{1680}\; R\, R_{\mu\nu\rho\sigma}^{2}-\frac{991}{5040} \; R_{\mu\nu\rho\sigma}R^{\rho\sigma\alpha\beta}R_{\alpha\beta}{}^{\mu\nu}-\frac{71}{180}\, R_{\alpha\mu\nu\beta}R^{\mu\rho\sigma\nu}R_{\rho}{}^{\alpha\beta}{}_{\sigma}\ . \label{a3D6}
\end{align}
Also, its expression on maximally symmetric spaces (MSS) may be useful for future reference. As we are not aware of any such calculation carried out in the literature we list it here. Using the relations in appendix \ref{appendixA}, we find that \eqref{a3} on MSS collapses into 
\begin{align} \label{94}
a_{3}^{\rm MSS}(D) = \frac{35 D^6-217 D^5-3257 D^4-9239D^3+37470D^2+183672 D-302400}{90720 (D-1)^2 D^2} \; R^3 \ ,
\end{align}
yielding in $D=6$
\begin{align}
\left.a_{3}^{\rm MSS}\right|_{D=6} =-\frac{3181}{63000}\; R^3\ .
\end{align}

%#####################################################
\subsubsection{Ghost and Graviton separately}
%#####################################################

For future comparison with the results of the heat kernel method calculation, it is useful to project on the degrees of freedom of the ghost and graviton respectively. To this task, we must use the right measures in the path integral \eqref{4}. These can be found generalizing to our case the worldline partition functions of \cite{Bastianelli:2013tsa} leading us to 
\begin{align}
&P_{\rm gh}(z,\omega)=\frac{(z+1)^D}{z^2}\frac{1}{\omega }\ , \\[.5em]
&P_{\rm gr}(z,\omega)=\frac{2 (z+1)^D}{\omega z^2}+\frac{(\omega +1)^{D-2} (z+1)^{D-2} (z-\omega )^2 (\omega z-1)}{2 \omega ^3 z^3}\ .
\end{align}
Indeed, note that
\begin{equation}\label{SeparateMeas}
P_{\rm gr}(z,\omega) -2 \, P_{\rm gh}(z,\omega)=P(z,\omega)\ .
\end{equation}
It is therefore possible to obtain the contributions to the Seeley-DeWitt coefficients coming only from the ghost (graviton) projecting onto the desired Hilbert space via $P_{\rm gh}\,(P_{\rm gr})$. Let us emphasize that these results are not new \emph{per se} since the heat kernel procedure requires calculating them individually and then putting them together, as we shall see. What is new here is the possibility of obtaining them also from the worldline viewpoint of the ${\cal N}=4$ particle. Therefore, regarding the ghost we have
\begin{align}
\begin{split}
\label{a0gh} a_{0}^{\rm gh}(D)&= D
\end{split}\\[.5em]
\begin{split}
\label{a1gh} a_{1}^{\rm gh}(D)&= \frac{D+6}{6}\; R
\end{split}\\[.5em]
\begin{split}
\label{a2gh} a_{2}^{\rm gh}(D)&=\frac{5 D^2+58 D+180}{360 D}\; R^2 + \frac{D-15}{180}\; R_{\mu\nu\rho\sigma}^{2}
\end{split}\\[.5em]
\begin{split}
\label{a3gh} a_{3}^{\rm gh}(D)&=\frac{35 D^3+588 D^2+3512 D+7560}{45360 D^2}\; R^3 +\frac{7 D^2-62 D-714}{7560 D}\;R\, R_{\mu\nu\alpha\beta}^{2} \\[.3em]
&\phantom{=}+\frac{17 D-252}{45360}\;R_{\mu\nu\rho\sigma}R^{\rho\sigma\alpha\beta}R_{\alpha\beta}{}^{\mu\nu} +\frac{D-18}{1620}\;R_{\alpha\mu\nu\beta}R^{\mu\rho\sigma\nu}R_{\rho}{}^{\alpha\beta}{}_{\sigma}\ ,
\end{split}
\end{align}
while concerning the graviton 
\begin{align}
\begin{split}
\label{a0gr} a_{0}^{\rm gr}(D)&=\frac{1}{2} D (D+1) 
\end{split}\\[.5em]
\begin{split}
\label{a1gr} a_{1}^{\rm gr}(D)&= \frac{1}{12} \left(D^2+D-12\right)\; R
\end{split}\\[.5em]
\begin{split}
\label{a2gr} a_{2}^{\rm gr}(D)&=\frac{\left(5 D^2+3 D-122\right)}{720} \; R^2 +\frac{\left(D^2-29 D+480\right)}{360} \; R_{\mu\nu\rho\sigma}^{2}
\end{split}\\[.5em]
\begin{split}
\label{a3gr} a_{3}^{\rm gr}(D)&= \frac{35 D^3-7 D^2-1318 D+488}{90720 D}\;R^3+ \frac{7 D^3-202 D^2+3109 D+9744}{15120 D}\;R\, R_{\mu\nu\alpha\beta}^{2} \\[.3em]
&\phantom{=}+\frac{17 D^2-487 D-16128}{90720}\;R_{\mu\nu\rho\sigma}R^{\rho\sigma\alpha\beta}R_{\alpha\beta}{}^{\mu\nu}+\frac{D^2-35 D-1152}{3240}\;R_{\alpha\mu\nu\beta}R^{\mu\rho\sigma\nu}R_{\rho}{}^{\alpha\beta}{}_{\sigma}\ .
\end{split}
\end{align}
One can easily see that the coefficient $a_0(D)$ correctly reproduces the expected degrees of freedom in $D$ dimensions. Moreover, it is immediate to check that by summing up the contributions as prescribed by \eqref{SeparateMeas} we obtain the correct total coefficients \eqref{a0}--\eqref{a3}.

%################################################################################################################
%################################################################################################################

%################################################################################################################
%################################################################################################################
\section{Heat kernel method} \label{sec3}

The gauge-invariant coefficients computed from the $\mathcal{N}=4$ spinning particle can be obtained in an equivalent but completely independent manner by exploiting the heat kernel method. This is a well-known technique from mathematical physics, which allows studying perturbatively second-order differential operators on Riemannian manifolds \cite{Avramidi2015}, and it has been applied extensively to quantum gravity starting from the work of DeWitt, \cite{DeWitt:1964mxt, DeWitt:1984sjp, DeWitt:2003pm}, as for example in \cite{Barvinsky:1985an, Fradkin:1985am, Avramidi:2000bm, Vassilevich:2003xt}. In this section, we briefly review the ideas behind this technique and apply it to the case of Euclidean perturbative quantum gravity, to reproduce the coefficients \eqref{a1}--\eqref{a3} and provide a strong consistency check for our computations.

%################################################################################################################
\subsection{Heat kernel and one-loop effective action}
%################################################################################################################

It is well known that the one-loop effective action for a bosonic or fermionic field theory living in a $D$-dimensional Euclidean space assumes the form (see for example \cite{Avramidi:2000bm})
\begin{equation}\label{Gamma1}
\Gamma_{(1)}[\Phi]=\frac{1}{2}\log{\text{sDet}{\Delta}}
= \frac{1}{2}{\text{sTr} \log {\Delta}}
\ ,
\end{equation}
where $\Delta$ is an elliptic second-order differential operator of the form
\begin{equation}\label{DiffOp}
\Delta= -\nabla_{\scriptscriptstyle \! \! ({\cal A})}^2 -V 
\end{equation}
while ``$\text{sDet}$'' is the Berezin functional superdeterminant and ``sTr'' the functional supertrace. The operator $\Delta$ is here taken to act on a scalar field $\phi$ which carries a representation of the (non-abelian) gauge field ${\cal A}_\mu$, contained in the connection that defines the full covariant derivative $\nabla^{\scriptscriptstyle ({\cal A})}_\mu = \nabla_\mu + {\cal A}_\mu$. It has an associated gauge field strength $\Omega_{\mu\nu}$, defined by the commutator of the covariant derivatives on the scalar field $\phi$, $ [\nabla^{\scriptscriptstyle ({\cal A})}_\mu, \nabla^{\scriptscriptstyle ({\cal A})}_\nu]\phi = \Omega_{\mu\nu}\phi$. The Laplacian is defined as usual by $\nabla_{\scriptscriptstyle \! \! ({\cal A})}^2 =g^{\mu\nu} \nabla^{\scriptscriptstyle ({\cal A})}_\mu\nabla^{\scriptscriptstyle ({\cal A})}_\nu$
and we consider a potential $V$ which is matrix valued, just like the gauge field ${\cal A}_\mu$. Thus, the elliptic second-order differential operator $\Delta$ depends on the metric $g_{\mu\nu}$
and on the matrix valued potential ${\cal A}_\mu$ and $V$. 

Using Schwinger proper-time parametrization, we can rewrite \eqref{Gamma1} as
\begin{equation}\label{Gamma1_2}
\Gamma_{(1)} = \frac{1}{2}{\text{sTr} \log {\Delta}} =-\frac{1}{2}\int_0^\infty \frac{\diff{T}}{T}\ \text{sTr} \exp{(-T\Delta)}\ ,
\end{equation}
where an infinite additive constant is neglected, as usual. The operator $\exp{(-T \Delta)}$ is known as the heat semigroup of the operator $\Delta$, and its integral kernel is the heat kernel $U(T; x,x')$ 
(corresponding to the matrix elements $\langle x| \exp{(-T \Delta)}| x' \rangle$ in quantum mechanical terms), which in the following will always be considered at coinciding points $x'\to x$. Plugging the explicit form of $\Delta$ given by \eqref{DiffOp} inside \eqref{Gamma1_2}, and introducing a mass term by shifting $\Delta \to \Delta +m^2$, so to have an infrared regulating mass if the original field was massless, we have
\begin{equation}\label{Gamma1bis}
\Gamma_{(1)}=-\frac{1}{2}\int_0^\infty \frac{\diff{T}}{T}\ \exp{(-T m^2)}\int \diff{^Dx}\ \sqrt{g}\ \text{str} \, {U(T; x,x)}\ .
\end{equation}
Note that the leftover supertrace ``str" is now to be performed only over the remaining discrete indices carried by the representation of the field $\phi$ on which $\Delta$ acts upon.

At this point, we employ the heat kernel expansion for small Euclidean time $T\to 0$, at coinciding points, which takes the form \cite{DeWitt:1964mxt, DeWitt:1984sjp, DeWitt:2003pm}
\begin{equation}\label{HDMS}
U(T;x,x)\sim (4\pi T)^{-\frac{D}{2}}\sum_{j=0}^{\infty} T^j a_j(x)\ ,
\end{equation}
where the heat kernel coefficients $a_j(x)$, also known as Seeley-DeWitt coefficients, can be expressed in terms of the metric and gauge invariants of the manifold. With the aid of \eqref{HDMS}, we are able to identify the divergences of the one-loop effective action \eqref{Gamma1bis} precisely with a subset of the Seeley-DeWitt coefficients, namely the ones that produce a divergence in the proper time integration at $T\to 0$ in
\begin{equation}\label{Gamma1ter}
\Gamma_{(1)}=-\frac{1}{2}\int_0^\infty \frac{\diff{T}}{T}\ \exp{(-T m^2)}\int \frac{\diff{^Dx}\,\sqrt{g}}{(4\pi T)^{\frac{D}{2}}}\ \sTr{\sum_{j=0}^{\infty} T^j a_j(x)}\ .
\end{equation}
For example, at $D=4$ the diverging terms are associated with the coefficients $a_0(x), a_1(x), a_2(x)$, while at $D=6$ also $a_3(x)$ leads to an additional divergence (the logarithmic divergence in that dimension).
The problem of finding the UV divergences of the effective action is then reduced to the computation of Seeley-DeWitt coefficients for a generic theory, which has already been carried out up to $a_5(x)$ \cite{Avramidi2015}. For application to perturbative quantum gravity, we will consider here the first four coefficients, i.e. from $a_0(x)$ to $a_3(x)$. The fourth coefficient $a_3(x)$, in particular, has been computed for the first time by Gilkey \cite{Gilkey:1975iq}, and later confirmed by Avramidi through a fully covariant method \cite{Avramidi:1990je}.\footnote{It is important to note that the notation we employ here for the Riemann tensor, Ricci tensor, and scalar is the same of \cite{Bastianelli:2022pqq, Bastianelli:2000hi}, see equation \eqref{Notation}, while \cite{Gilkey:1975iq, Vassilevich:2003xt} adopt the opposite sign in the Riemann tensor and in its contractions, $R_{\mu\nu}\equiv R_{\alpha\mu\nu}^{\ \ \ \ \alpha}$, so the heat kernel coefficients \eqref{HK0}--\eqref{HK3} are to be modified accordingly.} 

We now list the general results for the coefficients corresponding to the operator \eqref{DiffOp} up to $a_3(x)$, as taken from \cite{Gilkey:1975iq, Bastianelli:2000hi}. For the sake of brevity, it is useful to define as done in \cite{Bastianelli:2000hi}
\begin{equation}\label{HKalphabeta}
a_j(x) \equiv \alpha_j(x) + \beta_j(x)\ ,
\end{equation}
where the first term $\alpha_j(x)$ comes from considering an exponentiated form of the heat kernel series,
\begin{equation}\label{HKconnected}
\sTr{\left[\sum_{j=0}^{\infty} T^j a_j(x) \right]}\equiv \sTr{\left[\exp{\left(\sum_{j=1}^{\infty} T^j \alpha_j(x)\right)}\right]}\ ,
\end{equation}
while the second one $\beta_j(x)$ is the remainder, which, up to the third order, is evaluated as
\begin{equation}\label{BetaValues}
\beta_0=\beta_1 = 0\ ,\quad
\beta_2 = \frac{1}{2}\alpha_1^2\ ,\quad
\beta_3 = \frac{1}{6}\alpha_1^3+\alpha_1\alpha_2\ .
\end{equation}
The coefficients $\alpha_j(x)$ are given by 
\begin{align}
\begin{split}
\label{HK0} \alpha_0(x) &= \mathbbm{1}
\end{split}\\[.5em]
\begin{split}
\label{HK1} \alpha_1(x) &= \frac{1}{6}R\mathbbm{1} + V
\end{split}\\[.5em]
\begin{split}
\label{HK2} \alpha_2(x) &= \frac{1}{6}\nabla^2\left(\frac{1}{5}R\mathbbm{1}+V\right)+ \frac{1}{180}\left(R_{\mu\nu\rho\sigma}^2-R_{\mu\nu}^2\right)\mathbbm{1} + \frac{1}{12}\Omega_{\mu\nu}^2
\end{split}\\[.5em]
\begin{split} \label{HK3}
\alpha_3(x) &= \frac{1}{7!}\left[18\nabla^4 R + 17(\nabla_\mu R)^2-2(\nabla_\mu R_{\nu\sigma})^2 - 4\nabla_\mu R_{\nu\sigma}\nabla^\nu R^{\mu\sigma}+9(\nabla_\alpha R_{\mu\nu\rho\sigma})^2 -8R_{\mu\nu}\nabla^2 R^{\mu\nu}\right. \\
&\phantom{=}
+12R^{\mu\nu}\nabla_\mu \nabla_\nu R
+12R_{\mu\nu\rho\sigma}\nabla^2R^{\mu\nu\rho\sigma} 
+ \frac{8}{9}R_{\mu}{}^{\nu}R_{\nu}{}^{\sigma}R_{\sigma}{}^{\mu} - \frac{8}{3}R_{\mu\nu}R_{\rho\sigma}R^{\mu\rho\nu\sigma}
\\
&\phantom{=}\left.-\frac{16}{3}R_{\mu\nu}R^\mu{}_{\rho\sigma\tau}R^{\nu\rho\sigma\tau}+\frac{44}{9}R_{\mu\nu}{}^{\rho\sigma}R_{\rho\sigma}{}^{\alpha\beta}R_{\alpha\beta}{}^{\mu\nu}+\frac{80}{9}R_{\mu\nu\rho\sigma}R^{\mu\alpha\rho\beta}R^{\nu}{}_{\alpha}{}^{\sigma}{}_{\beta}\right]\mathbbm{1}\\
&\phantom{=}+\frac{2}{6!}\left[8(\nabla_\mu \Omega_{\nu\sigma})^2+2(\nabla^\mu \Omega_{\mu\nu})^2 + 12\Omega_{\mu\nu}\nabla^2\Omega^{\mu\nu}-12\Omega_{\mu}{}^{\nu}\Omega_{\nu}{}^{\sigma}\Omega_{\sigma}{}^{\mu}+6R_{\mu\nu\rho\sigma}\Omega^{\mu\nu}\Omega^{\rho\sigma}-4R_{\mu\nu}\Omega^{\mu\sigma}\Omega^\nu{}_{\sigma}\right.\\
&\phantom{=}\left. +6\nabla^4 V+30(\nabla_\mu V)^2+4R_{\mu\nu}\nabla^\mu\nabla^\nu V+12\nabla_\mu R\nabla^\mu V\right]\ .
\end{split}
\end{align}
They will be used in the next section.

%################################################################################################################
\subsection{Euclidean quantum gravity}
%################################################################################################################

Consider a $D$-dimensional Riemannian manifold $(\mathcal{M}, \bm{G})$ equipped with a metric tensor $\bm{G}$ having Euclidean signature. The starting point for our treatment of gravity is the Einstein-Hilbert action,
\begin{equation}\label{EH}
S_{\rm EH}[\bm{G}]=-\frac{1}{k^2}\int\diff{^D x}\ \sqrt{G}\left[R(\bm{G})-\Lambda\right]\ ,
\end{equation}
where $k^2 \equiv 16\pi G_{\rm N}$, being $G_{\rm N}$ the Newton constant, $R(\bm{G})$ is the Ricci scalar computed from $\bm{G}$, and $G\equiv \left|\det{G_{\mu\nu}}\right|$. A cosmological constant $\Lambda$ has also been included. By employing the background field method, we split the metric tensor $\bm{G}$ into a fixed classical background $\bm{g}$ and ``small'' quantum perturbations $\bm{h}$, namely:
\begin{equation}\label{Splitting}
G_{\mu\nu}(x) = g_{\mu\nu}(x) + h_{\mu\nu}(x)\ .
\end{equation}
As a consequence of this splitting, the action \eqref{EH} can be expanded in power series in the fluctuations $\bm{h}$. Since we are interested in the one-loop level of accuracy, we will be concerned with the second-order term in $\bm{h}$, which reads 
\begin{align}
\label{S2} S_2 &= \int \diff{^Dx} \sqrt{g} \left[-\frac{1}{4}h^{\mu\nu}\left(\nabla^2+2\Lambda-R\right)h_{\mu\nu} +\frac{1}{8}h\left(\nabla^2+2\Lambda-R\right) -\frac{1}{2}\left(\nabla^\nu h_{\mu\nu}-\frac{1}{2}\nabla_\mu h\right)^2\right. \nonumber \\
&\phantom{= \int \diff{^Dx} \sqrt{g}+}\left. - \frac{1}{2}\left(h^{\mu\lambda}h_\lambda{}^{\nu} - hh^{\mu\nu}\right)R_{\mu\nu} -\frac{1}{2}h^{\mu\lambda}h^{\nu\rho}R_{\mu\nu\lambda\rho}\right]\ .
\end{align}
It is important to note that in \eqref{S2} the Ricci tensor $R_{\mu\nu}=R_{\mu\nu}(\bm{g})$ and scalar $R=R(\bm{g})$, as well as covariant derivatives $\nabla_\mu = \nabla_\mu(\bm{g})$, are computed with respect to the background metric $\bm{g}$. The gauge symmetries acting on $\bm{h}$ and leaving the background metric $\bm{g}$ invariant are then BRST quantized by introducing the ghost $c$ and antighost $\overline{c}$ fields, and adding to the action the Slavnov variation of de Donder gauge-fixing function $f_\mu =\nabla^\nu h_{\mu\nu}-\frac{1}{2}\nabla_\mu h$, see for example \cite{Bastianelli:2013tsa} for further details. The final result is
\begin{equation}
S_2[\bm{h},c,\overline{c}] = S_{\rm gr}[\bm{h}]+S_{\rm gh}[c,\overline{c}]\ ,
\end{equation}
where
\begin{align}
\begin{split}
S_{\rm gr}[\bm{h}] &= \int \diff{^Dx} \sqrt{g} \left[-\frac{1}{4}h^{\mu\nu}\left(\nabla^2+2\Lambda-R\right)h_{\mu\nu} +\frac{1}{8}h\left(\nabla^2+2\Lambda-R\right)\right. \\
&\phantom{= \int \diff{^Dx} \sqrt{g}+}\left. - \frac{1}{2}\left(h^{\mu\lambda}h_\lambda{}^{\nu} - hh^{\mu\nu}\right)R_{\mu\nu} -\frac{1}{2}h^{\mu\lambda}h^{\nu\rho}R_{\mu\nu\lambda\rho}\right]\ ,
\end{split}\\ 
\begin{split}
S_{\rm gh}[c,\overline{c}] &= \int \diff{^Dx} \sqrt{g}\ \bar{c}^\mu \left(\nabla^2 c_\mu + R_{\mu\nu}c^\nu\right)\ .
\end{split}
\end{align}
We are now able to identify from the actions $S_{\rm gr}$ and $S_{\rm gh}$ the invertible kinetic operators for the graviton and ghost fields, denoted by $F_{\mu\nu\alpha\beta}$ and $\mathcal{F}_{\mu\nu}$, respectively. By setting
\begin{equation}
S_{\rm gr}[\bm{h}] =\int \diff{^Dx}\ \sqrt{g}\ \frac{1}{2}h_{\mu\nu}F^{\mu\nu\alpha\beta}h_{\alpha\beta} \qquad\mbox{and}\qquad S_{\rm gh}[c,\overline{c}]=\int \diff{^Dx}\ \sqrt{g}\ \bar{c}_\mu \mathcal{F}^{\mu}{}_{\nu} c^\nu\ ,
\end{equation}
and exploiting the properties of Einstein spaces, we find
\begin{align}
\label{GravitonKin} F_{\mu\nu}{}^{\alpha\beta} &= -\frac{1}{2}\left(\delta_\mu^\alpha\delta_\nu^\beta+\delta_\mu^\beta\delta_\nu^\alpha\right)\nabla^2 - R_{\mu}{}^{\alpha}{}_{\nu}{}^{\beta} - R_{\mu}{}^{\beta}{}_{\nu}{}^{\alpha}\\[.3em]
\label{GhostKin} \mathcal{F}^{\mu}_{\ \nu} &= \delta^\mu_\nu \left(\nabla^2+\frac{1}{D}R\right)\ ,
\end{align}
where the graviton operator indices are raised and lowered with the DeWitt supermetric
\begin{equation}\label{DeWittSM}
\gamma^{\mu\nu\alpha\beta} \equiv \frac{1}{4}\left(g^{\mu\alpha}g^{\nu\beta}+g^{\mu\beta}g^{\nu\alpha}-g^{\mu\nu}g^{\alpha\beta}\right)\ ,\quad
\gamma_{\mu\nu\alpha\beta} = g_{\mu\alpha}g_{\nu\beta}+ g_{\mu\beta}g_{\nu\alpha} - \frac{2}{D-2}g_{\mu\nu}g_{\alpha\beta}\ .
\end{equation}

The reasons for immediately reducing to Einstein spaces --- that is, to compute the coefficients directly on-shell --- are twofold. Firstly, this allows us a direct comparison with the worldline results, even separately for ghost and graviton, since the on-shell condition is forced by the quantum consistency of the $\mathcal{N}=4$ spinning particle. Secondly, proceeding otherwise the results would not be gauge-invariant, but rather would depend on the gauge chosen, as expected for gauge theories, see for instance the recent analysis carried out in \cite{Brandt:2022und}.

Since \eqref{GravitonKin}--\eqref{GhostKin} are elliptic second-order differential operators, they can be treated within the heat kernel expansion, and the coefficients \eqref{HK0}--\eqref{HK3} can be computed by identifying the explicit formulae for $\mathbbm{1}$, $V$ and $\Omega_{\mu\nu}$. The ghost field configuration space is $D$-dimensional, and by comparing \eqref{GhostKin} with \eqref{DiffOp}, as well as recalling that
\begin{equation}\label{GhostComm}
[\nabla_\mu, \nabla_\nu]c^\rho=R_{\mu\nu}{}^{\rho}{}_{\sigma}c^\sigma\ ,
\end{equation}
we conclude that the substitutions to be performed in the heat kernel coefficients \eqref{HK0}--\eqref{HK3} are
\begin{equation}\label{GhostSost}
\begin{cases}
\mathbbm{1}\ \leftrightarrow\ \delta^\mu_\nu\\[.3em]
V\ \leftrightarrow\ \dfrac{1}{D}R\delta^\mu_\nu\\[.5em]
(\Omega_{\mu\nu})^\rho_{\ \sigma} \ \leftrightarrow\ R_{\mu\nu}{}^{\rho}{}_{\sigma}\ .
\end{cases}
\end{equation}
Note that in this expression the indices $\mu$, $\nu$ label the different elements of the gauge field strength $\Omega_{\mu\nu}$, which are $D\times D$ matrices whose components are given by the (spacetime) indices $\rho$, $\sigma$. On the other hand, for the graviton field the configuration space is $\frac{1}{2}D(D+1)$ dimensional (space of symmetric tensors) and the substitutions to be performed are
\begin{equation}\label{GravitonSost}
\begin{cases}
\mathbbm{1}\ \leftrightarrow\ \delta_{\mu\nu}^{\ \ \ \alpha\beta}\\
V\ \leftrightarrow\ \mathcal{V}_{\mu\nu}{}^{\alpha\beta}\\
(\Omega_{\mu\nu})_{\rho\sigma}{}^{\alpha\beta} \ \leftrightarrow\ R_{\rho\sigma}{}^{\alpha\beta}{}_{\mu\nu}
\end{cases}
\end{equation}
where
\begin{align}
\label{SuperDelta} \delta_{\mu\nu}{}^{\alpha\beta} &\equiv \frac{1}{2}\left(\delta_\mu^\alpha\delta_\nu^\beta+\delta_\mu^\beta\delta_\nu^\alpha\right)\\[.5em]
\label{VGraviton} \mathcal{V}_{\mu\nu}{}^{\alpha\beta} &\equiv R_{\mu}{}^{\alpha}{}_{\nu}{}^{\beta} + R_{\mu}{}^{\beta}{}_{\nu}{}^{\alpha}\ ,
\end{align}
and the commutator is given by a symmetrized version of the Riemann tensor
\begin{equation}\label{GravitonComm}
[\nabla_\mu, \nabla_\nu]h_{\rho\sigma}=R_{\rho\sigma}{}^{\alpha\beta}{}_{\mu\nu}\,h_{\alpha\beta}\ ,
\end{equation}
where
\begin{equation}\label{SuperRiemann}
R_{\rho\sigma}{}^{\alpha\beta}{}_{\mu\nu}\equiv \frac{1}{2}\left(\delta_\rho^\alpha R_{\sigma}{}^{\beta}{}_{\mu\nu} + \delta_\rho^\beta R_{\sigma}{}^{\alpha}{}_{\mu\nu} +\delta_\sigma^\alpha R_{\rho}{}^{\beta}{}_{\mu\nu} + \delta_\sigma^\beta R_{\rho}{}^{\alpha}{}_{\mu\nu} \right)\ .
\end{equation}
At this point, the computations are tedious but straightforward, see appendix \ref{appendixC} for details. The final results for the ghost and graviton fields separately are precisely the coefficients \eqref{a0gh}--\eqref{a3gh} and \eqref{a0gr}--\eqref{a3gr} obtained from the worldline formalism. The total coefficients for the physical graviton, according to the supertrace appearing in \eqref{Gamma1ter}, as recognized also 
from \eqref{SeparateMeas}, are given by
\begin{equation}\label{TotHKgen}
\Tr{\left[a_j\right]}=\Tr{\left[a_j^{\rm gr}\right]}-2\Tr{\left[a_j^{\rm gh}\right]}\ .
\end{equation}
Again, the results obtained reproduce the ones coming from worldline computations \eqref{a1}--\eqref{a3}, providing a strong cross-check for the correctness of both.

%################################################################################################################
%################################################################################################################

%################################################################################################################
%################################################################################################################
\section{On one-loop divergences of quantum gravity} \label{sec4}

The coefficients \eqref{a1}--\eqref{a3}, which include the newly computed coefficient $a_3(D)$, allow for further investigations of the issue of divergences in the quantum theory of gravity. Thus, let us review some crucial results from the literature and discuss how our newly-calculated coefficient $a_3(D)$ fits into the picture, thus providing us with additional confirmation of the validity of our result. We focus our discussion on the spacetime of dimensions $D=4$ and $D=6$, as in these cases there are no additional divergences on top of the one we have already computed (new divergences start to appear from $D=8$ onwards). The type of divergences arising in quantum gravity emerge naturally from the representation of the one-loop effective action with a short proper time expansion, which we can read both from the worldline viewpoint \eqref{series} and from the heat kernel one \eqref{Gamma1ter}:
\begin{equation}
\Gamma[g_{\mu\nu}]= -\frac{1}{2}\int_{0}^{\infty}\frac{dT}{T^{1+\frac{D}{2}}}\int \frac{d^{D}x_{0}}{\left(4\pi \right)^{\frac{D}{2}}}\sqrt{g(x_{0})}\left[a_0 + a_1 T + a_2 T^2 + a_3 T^3+ \mathcal{O}(T^{4})\right]\ . \label{divergences}
\end{equation} 
We are interested in studying the UV divergences that arise from the $T\rightarrow0$ limit of the proper time integration. Setting $D=4$ we recognize that possible divergences arise from the coefficients $a_0, a_1,a_2$,
with $a_2$ being associated with the logarithmic divergence. In $D=6$, also $a_3$ gives rise to an additional divergence, the logarithmic one in that dimension.

One may wonder how to relate the $\frac{1}{\epsilon}$ pole of dimensional regularization in QFT, widely present in the literature, with our situation. To address this point, it is useful to evaluate the proper time integral term by term in \eqref{divergences}, to display the gamma function dependence. We find 
\begin{equation}
\int_{0}^{\infty}\frac{dT}{T^{1+\frac{D}{2}}}T^{p} \, e^{-m^{2}T}= (m^{2})^{\frac{D}{2}-p}\, \Gamma\left(p-\frac{D}{2}\right)\ ,
\end{equation}
where $p=2,3$ correspond to our cases of interest $D=4,6$, respectively. Now, using dimensional regularization, namely taking $D=2p-2\epsilon$ and expanding the gamma function, we see the appearance of the usual $\frac{1}{\epsilon}$ pole as the leading divergent term: it corresponds precisely to the logarithmic divergences seen in dimensional regularization \cite{Schwartz:2014sze}. 

In general, one has to deal also with IR divergences: for the sake of our discussion, they can be avoided either by introducing an upper cutoff in the proper time or by keeping a ``small'' mass regulator $m$, as we have done above.
 
%################################################################################################################
\subsection{Pure gravity in four dimensions}
%################################################################################################################

It has long been known since the pioneering work of 't Hooft and Veltman \cite{tHooft:1974toh}, that pure gravity with vanishing cosmological constant is a renormalizable theory at one-loop in $D=4$. It is free of logarithmic divergences, while other divergences are not seen in dimensional regularization, and in any case they can be eliminated by renormalization. The same does not hold in the case of a non-vanishing cosmological constant, as found by Christensen and Duff \cite{Christensen:1979iy}. Let us briefly review these statements in the light of our calculations. Setting $D=4$ in \eqref{divergences} we see that the different powers of $T$ give rise to the quartic, quadratic, and logarithmic divergences parametrized by $a_0$, $a_1$ and $a_2$, respectively. In dimensional regularization only the logarithmic divergences are visible. From \eqref{a2} we read
\begin{equation}
\left. a_{2}\right|_{D=4}=-\frac{29}{40}\, R^2+\frac{53}{45} \, R_{\mu\nu\rho\sigma}^{2}\ .
\end{equation}
These numerical values for the one-loop four-dimensional logarithmic divergences of quantum gravity with non-vanishing cosmological constant coincide precisely with those calculated long ago by Christensen and Duff.\footnote{For comparison, one has to make evident the cosmological constant term with the on-shell condition \eqref{1EM}, which reads $R=4\Lambda$.} The term proportional to $R_{\mu\nu\rho\sigma}^{2}$ could be neglected, as thanks to the Gauss-Bonnet theorem for four-dimensional Einstein manifolds it is proportional to a total derivative, and thus eliminable from the effective action, but the remaining term proportional
to $R^2$ cannot be renormalized away by redefining the parameters of the Einstein-Hilbert action. The theory is not renormalizable.
 
On the other hand, setting the cosmological constant to vanish, one finds that the on-shell background satisfies $R_{\mu\nu}=0$, and thus $R=0$. The logarithmic divergence reduces to 
\begin{equation}
a_{2}\Big|_{\substack{D=4\\[0.1em] \Lambda=0}} = \frac{53}{45} \, R_{\mu\nu\rho\sigma}^{2} 
\end{equation}
which in four dimensions is a total derivative, as discussed earlier, and can be eliminated from the effective action. Thus, one recovers the result that the one-loop logarithmic divergences of pure quantum gravity without cosmological constant vanish in four dimensions. This property does not hold true anymore at two-loops, as found by Goroff and Sagnotti \cite{Goroff:1985th} and verified by van de Ven \cite{vandeVen:1991gw}.
Returning to the one-loop divergences for vanishing cosmological constant, one finds that also $a_1$ vanishes. This leaves only the quartic divergence proportional to $a_0$, which gives the number of degrees of freedom of the graviton. It requires a renormalization of the cosmological constant back to zero, which makes the theory rather unnatural in the technical sense of 't Hooft \cite{tHooft:1979rat}, but in any case renormalizable at one-loop.

For arbitrary nonzero values of the cosmological constant, the quadratic divergence related to $a_1$ is not vanishing anymore and its value at $D=4$ 
\begin{equation}
\left. a_{1}\right|_{D=4}= -\frac83\, R
\end{equation}
reproduces the gauge-invariant result already computed in \cite{Bastianelli:2013tsa, Martini:2021slj}. It can be renormalized away by redefining the Newton constant. Finally, the coefficient $a_3$ gives rise to a finite term in the four-dimensional effective action, but its physical meaning is unclear. It is gauge invariant, but infrared divergences invalidate a local expansion of the effective action as delivered by the small proper time approximation of the heat kernel, which is only useful to locate the UV divergences, as far as we know. It might however signal some property of quantum gravity which we are unaware of.

%################################################################################################################
\subsection{Pure gravity in six dimensions}
%################################################################################################################

The newly computed coefficient $a_3$ \eqref{a3}, allows us to see what happens in six spacetime dimensions. Setting $D=6$ in the coefficients \eqref{a0}--\eqref{a3} we find
\begin{align}
\begin{split} 
\left. a_{0} \right|_{D=6}
&= 9 
\;, \qquad \qquad \quad
\left. a_{1}\right|_{D=6} = -\frac32 \; R
\;, \qquad \qquad \quad
\left. a_{2} \right|_{D=6}= -\frac{11}{20}\; R^2 +\frac{21}{20} \; R_{\mu\nu\rho\sigma}^{2}
\end{split}\\[.5em]
\begin{split}
\left. a_{3}\right|_{D=6} &= -\frac{799}{11340} \;R^3
+\frac{481}{1680} \;R\,R_{\mu\nu\rho\sigma}^{2}
-\frac{991}{5040} \;R_{\mu\nu\rho\sigma}R^{\rho\sigma\alpha\beta}R_{\alpha\beta}{}^{\mu\nu}
-\frac{71}{180}\; R_{\alpha\mu\nu\beta}R^{\mu\rho\sigma\nu}R_{\rho}{}^{\alpha\beta}{}_{\sigma}\ ,
\end{split}
\end{align}
that furnish the full list of one-loop divergences of quantum gravity with cosmological constant in six dimensions. We stress that these coefficients are gauge invariant, and thus any other method of calculation should reproduce these values. 

A comparison with the literature can be made by setting the cosmological constant to zero and considering the logarithmic divergence parametrized by $a_3$, which reduces to
\begin{equation} \label{a3D6v2}
a_{3}\Big|_{\substack{D=6\\[0.1em] \Lambda=0}}=-\frac{991}{5040} \; R_{\mu\nu\rho\sigma}R^{\rho\sigma\alpha\beta}R_{\alpha\beta}{}^{\mu\nu} -\frac{71}{180}\, R_{\alpha\mu\nu\beta}R^{\mu\rho\sigma\nu}R_{\rho}{}^{\alpha\beta}{}_{\sigma}\ .
\end{equation}
These two remaining terms are proportional to two invariants that are generally independent of each other. However, it turns out that in six dimensions there exists an integral relation that connects them. It involves the use of the Gauss-Bonnet theorem and the introduction of the Euler character $\chi_{\rm E}(\mathcal{M})$, as explained in \cite{vanNieuwenhuizen:1976vb} and discussed more extensively in appendix \ref{appendixD}. Bottom line, we can further simplify the coefficient $a_3$, which becomes 
\begin{equation} \label{DivD6}
a_{3}\Big|_{\substack{D=6\\[0.1em] \Lambda=0}}=-\frac{9}{15120} \; R_{\mu\nu\rho\sigma}R^{\rho\sigma\alpha\beta}R_{\alpha\beta}{}^{\mu\nu}\; .
\end{equation}
It encodes the one-loop logarithmic divergences of pure gravity in six dimensions. We are now in the position of carrying out a comparison with the literature: our value in \eqref{DivD6} is in complete agreement with van Nieuwenhuizen's pioneering calculation \cite{VanNieuwenhuizen:1977ca}, besides a computational error in the numerical factor in his equation (81), already noted a year later by Critchley \cite{Critchley:1978kb}. 
Furthermore, confirmation of this value is also found in more recent works, see for instance \cite{Gibbons:1999qz, Dunbar:2002gu}. 

As for the general case of arbitrary cosmological constant, we are not aware of similar calculations, although they would certainly be interesting to pursue to further verify our findings.

%################################################################################################################
%################################################################################################################
\section{Conclusions} \label{sec5}

In this work, we have investigated the computation of counterterms necessary for the renormalization of the one-loop effective action of quantum gravity with cosmological constant in arbitrary dimensions. Our results are complete for dimensions $D<8$. The counterterms have been computed on-shell, so they furnish gauge invariant quantities characteristic of the quantum theory of gravity. 

Our main contribution was the determination of the Seleey-DeWitt coefficient $a_3(D)$ of perturbative quantum gravity, which to our knowledge has never been reported in its full generality in the literature. When restricted to six dimensions, it parameterizes the logarithmic divergence which was previously known only for the case of vanishing cosmological constant. To cross-check our calculations, we have used two distinct methods, a first-quantized description of the graviton in terms of the ${\cal N}=4$ spinning particle and the time-honored heat kernel method, finding complete agreement.
 
While the utility of heat kernel methods is well-known and they keep being employed in many contexts, see for example \cite{Bastianelli:2019zrq, Casarin:2023ifl} for some recent applications to trace anomalies,
first-quantized methods for treating the graviton with the ${\cal N}=4$ spinning particle are more recent and we have championed them here to show their usefulness. In this respect, it would be interesting to extend the present analysis to the first quantized model that describes the ${\cal N}=0$ supergravity \cite{Bonezzi:2020jjq}, i.e. the particle theory that has in its spectrum the graviton, the dilaton, and the antisymmetric tensor $B_{\mu\nu}$, as well as extend the present methods to the U($N$) spinning particles \cite{Marcus:1994em, Bastianelli:2009vj, Bastianelli:2011pe, Bastianelli:2012nh} to find a useful first-quantized way of describing gravitational theories on complex (K\"ahler) manifolds, and finally also consider double copy features on the worldline \cite{Bastianelli:2021rbt} to address gravitational aspects from a different perspective.

%################################################################################################################
%################################################################################################################

%################################################################################################################
%################################################################################################################
\section*{Acknowledgments}
%################################################################################################################
%################################################################################################################

F.F. would like to thank the Galileo Galilei Institute (Firenze, Italy) for its hospitality during part of this work. Our diagrams have been produced with the help of TikZ-Feynman~\cite{Ellis_2017}.

%################################################################################################################
%################################################################################################################
\appendix 
%################################################################################################################
%################################################################################################################

%################################################################################################################
%################################################################################################################
\section{Basis of invariants on Einstein manifolds}\label{appendixA}

We use the following conventions for the curvature tensors:
\begin{equation}\label{Notation}
[\nabla_\mu, \nabla_\nu] V^\lambda = 
R_{\mu\nu}{}^\lambda{}_\rho V^\rho \ , \ \ \ 
R_{\mu\nu}= R_{\lambda\mu}{}^\lambda{}_\nu 
\ , \ \ \ R= R^\mu{}_\mu > 0 \ {\rm on\ spheres.} 
\end{equation}
A $D$-dimensional Riemannian manifold without boundaries can be described through an (infinite) basis of curvature monomials $\mathcal{K}_i^n$. These are geometric invariants of order $n$ in the Riemann tensor, Ricci tensor, and scalar curvature, with two covariant derivatives counting as a Riemann tensor. They have been introduced by \cite{Fulling1992} and recently reviewed in \cite{Decanini2007, Decanini2008}. 
 At order $n=3$ we use the basis considered in \cite{Bastianelli:2000rs} which is made of $17$ independent invariants:
\begin{align}
\begin{array}{lll}
\mathcal{K}_1 = R^3 & 
\mathcal{K}_2 = R R_{\mu\nu}^2 & 
\mathcal{K}_3 = R R_{\mu\nu\rho\sigma}^2 \\ [3mm] 
\mathcal{K}_4 = R_\mu{}^\rho R_\rho{}^\nu R_\nu{}^\mu & 
\mathcal{K}_5 = R_{\mu\nu} R_{\rho\sigma} R^{\rho\mu\nu\sigma} & 
\mathcal{K}_6= R_{\mu\nu} R^{\mu\rho\sigma\lambda} R^\nu{}_{\rho\sigma\lambda} \\ [3mm] 
\mathcal{K}_7 = R_{\mu\nu}{}^{\rho\sigma} R_{\rho\sigma}{}^{\alpha\beta}R_{\alpha\beta}{}^{\mu\nu} \ \ \ \ &
\mathcal{K}_8 = R_{\mu\rho\sigma\nu} R^{\rho\alpha\beta\sigma} R_{\alpha}{}^{\mu\nu}{}_\beta \ \ \ \ &
\mathcal{K}_9 = R\nabla^2 R \\ [3mm]
\mathcal{K}_{10} = R_{\mu\nu}\nabla^2 R^{\mu\nu} &
\mathcal{K}_{11} = R_{\mu\nu\rho\sigma}\nabla^2 R^{\mu\nu\rho\sigma} &
\mathcal{K}_{12} = R^{\mu\nu} \nabla_\mu \nabla_\nu R \\ [3mm]
\mathcal{K}_{13} = (\nabla_\mu R_{\rho\sigma})^2 &
\mathcal{K}_{14} = \nabla_\mu R_{\nu\rho} \nabla^\nu R^{\mu\rho} &
\mathcal{K}_{15} = (\nabla_\alpha R_{\mu\nu\rho\sigma})^2 \\ [3mm]
\mathcal{K}_{16} = \nabla^2 R^2 &
\mathcal{K}_{17} =\nabla^4 R\ . &
\label{R3}
\end{array}
\end{align}
All other terms cubic in the curvature are linear combinations of the above invariants after taking into account the symmetry properties and the Bianchi identities of the Riemann tensor. 

On Einstein manifolds, the basis \eqref{R3} can be reduced further. Einstein metrics are defined by the equation
\begin{equation}
R_{\mu\nu}=\lambda g_{\mu\nu} 
\label{def-Em}
 \end{equation}
that upon contraction leads to $ R=\lambda D$. From the second Bianchi identity one finds that $R$ and $\lambda$ are constant for $D> 2$
\begin{equation}
\nabla^{\mu}R_{\mu\nu}=\frac{1}{2}\nabla_{\nu}R \quad\longrightarrow\quad (D-2) \nabla_{\nu}\lambda=0 \quad\longrightarrow\quad \nabla_{\nu}R=0 
\end{equation}
and then from \eqref{def-Em}, one finds that the Ricci tensor is also covariantly constant for $D> 2$
\begin{equation} \label{2EM}
\nabla_{\alpha}R_{\mu\nu}=0\, .
\end{equation}
Let us consider now the second Bianchi identity, namely
\begin{equation}
\nabla_{\rho}R_{\mu\nu\alpha\beta}+\nabla_{\beta}R_{\mu\nu\rho\alpha}+\nabla_{\alpha}R_{\mu\nu\beta\rho} = 0 \, ,
\end{equation}
by contracting the above identity with $g^{\rho\mu}$ and using \eqref{2EM} one gets 
\begin{equation} \label{div}
\nabla^{\mu}R_{\mu\nu\alpha\beta} = 0\ .
\end{equation}
We are now in the position to reduce the six-dimensional basis of invariants \eqref{R3} on Einstein manifolds to a minimal set of independent ones, namely
\begin{equation} \label{E3}
\mathcal{E}_1=R^{3} \;,
\hskip1cm \mathcal{E}_2= R\,R_{\mu\nu\rho\sigma}^{2}\;,
\hskip1cm \mathcal{E}_3=R_{\mu\nu\rho\sigma}R^{\rho\sigma\alpha\beta}R_{\alpha\beta}{}^{\mu\nu}\;,
\hskip1cm \mathcal{E}_4=R_{\alpha\mu\nu\beta}R^{\mu\rho\sigma\nu}R_{\rho}{}^{\alpha\beta}{}_{\sigma}\;.
\end{equation}
Indeed we have
\begin{equation}\label{EinsteinCond}
\mathcal{K}_9 = \mathcal{K}_{10} = 0\ , \qquad \mathcal{K}_2 = \frac{1}{D}\mathcal{E}_1\ ,\qquad \mathcal{K}_4 = -\mathcal{K}_5 = \frac{1}{D^2}\mathcal{E}_1\ ,\qquad \mathcal{K}_6 = \frac{1}{D}\mathcal{E}_2\ .
\end{equation}
Moreover, the only term of \eqref{R3} containing covariant derivatives and non-vanishing on Einstein manifolds, i.e. $\mathcal{K}_{11}$,\footnote{Since, up to a total derivative term, we have $\mathcal{K}_{15}=-\mathcal{K}_{11}$.} can be written as
\begin{align}
R_{\mu\nu\alpha\beta}\nabla^{2}R^{\mu\nu\alpha\beta}&=-R_{\mu\nu\alpha\beta}\nabla_{\rho}\left(
\nabla^{\beta}R^{\mu\nu\rho\alpha}+\nabla^{\alpha}R^{\mu\nu\beta\rho}
\right)=-2R_{\mu\nu\alpha\beta}\nabla_{\rho}\nabla^{\beta}R^{\mu\nu\rho\alpha}\nonumber\\
&=-2R_{\mu\nu\alpha\beta}\left(
\nabla^{\beta}\nabla_{\rho}R^{\rho\alpha\mu\nu}+R^{\beta}{}_{\lambda}R^{\lambda\alpha\mu\nu}+R_{\rho}{}^{\beta\alpha}{}_{\lambda}R^{\rho\lambda\mu\nu}+R_{\rho}{}^{\beta\mu}{}_{\lambda}R^{\rho\alpha\lambda\nu}+R_{\rho}{}^{\beta\nu}{}_{\lambda}R^{\rho\alpha\mu\lambda}
\right)\nonumber\\
&=-2R_{\mu\nu\alpha\beta}R_{\rho}{}^{\beta\alpha}{}_{\lambda}R^{\rho\lambda\mu\nu}+
\frac{2}{D}R\, R_{\mu\nu\alpha\lambda}^{2}
-4R_{\mu\nu\alpha\beta}R_{\rho}{}^{\beta\mu}{}_{\lambda}R^{\rho\alpha\lambda\nu}\nonumber\\
&=-R_{\mu\nu\alpha\beta}R^{\alpha\beta}{}_{\rho\lambda}R^{\rho\lambda\mu\nu}+\frac{2}{D}RR_{\mu\nu\alpha\beta}^{2}+4R_{\mu\nu\alpha\beta}R^{\nu\lambda\rho\alpha}R_{\lambda}{}^{\mu\beta}{}_{\rho}\nonumber\\
&=\frac{2}{D}\mathcal{E}_2 -\mathcal{E}_3 + 4\mathcal{E}_4\ ,
\end{align} 
where we made use of the second Bianchi identity, antisymmetry of the Riemann tensor and 
\begin{equation}
[\nabla_{\alpha},\nabla_{\beta}]R_{\mu\nu\rho\sigma}=R_{\alpha\beta \mu}{}^{\lambda}R_{\lambda\nu\rho\sigma}+R_{\alpha\beta\nu}{}^{\lambda}R_{\mu\lambda\rho\sigma}+R_{\alpha\beta\rho}{}^{\lambda}R_{\mu\nu\lambda\sigma}+R_{\alpha\beta \sigma}{}^{\lambda}R_{\mu\nu\rho\lambda}\ .
\end{equation}

Finally, let us discuss how the basis \eqref{E3} further simplifies on maximally symmetric spaces (MSS), which form a subset of the Einstein ones, where the Riemann tensor is given by
\begin{equation}
R_{\mu\nu\alpha\beta} = \frac{R}{D(D-1)}\left(g_{\mu\alpha}g_{\nu\beta}-g_{\mu\beta}g_{\nu\alpha}\right)\ .
\end{equation}
The invariants previously defined in \eqref{E3} all collaps to $\mathcal{E}_{1}$, and one finds
\begin{equation}
\mathcal{E}_{2} = \frac{2}{D(D-1)}\mathcal{E}_{1}\ ,\qquad \mathcal{E}_{3}=\frac{4}{D^{2}(D-1)^{2}}\mathcal{E}_{1}\ , \qquad \mathcal{E}_{4}=-\frac{D-2}{D^{2}(D-1)^{2}}\mathcal{E}_{1}\ .
\end{equation}
These relations allow us to evaluate the newly-computed coefficient $a_3(D)$ on MSS, giving \eqref{94} as a result.

%################################################################################################################
%################################################################################################################
\section{Worldline computations}\label{appendixB}

%################################################################################################################
\subsection{Worldline propagators}\label{appendixB1}
%################################################################################################################

The worldline propagators for the $\mathcal{N}=4$ spinning particle descend from the free action \eqref{freeaction}. Regarding the bosonic quantum fluctuations $q^{\mu}(\tau)$, we considered DBC during the evaluation of the derivative expansion of the effective action, namely $q^\mu(0)=q^\mu(1)=0$. This leads to the DBC worldline propagator defined by the two-point function 
\begin{equation} \label{A1}
\langle q^{\mu}(\tau)q^{\nu}(\sigma)\rangle = -2 Tg^{\mu\nu}(x_{0})\del_{{\rm D}}(\tau,\sigma)\ ,
\end{equation}
where
\begin{equation} \label{DBC}
\del_{{\rm D}}(\tau,\sigma)= \left(\tau-1\right)\sigma\, \theta\left(\tau-\sigma \right)+\left(\sigma-1\right)\tau\,\theta\left(\sigma-\tau \right)\;.
\end{equation}
We also list the derivatives of the DBC propagator
\begin{align}
\ddel_{\rm D}(\tau,\sigma)&=\sigma-\theta(\sigma-\tau)\label{derivata}\\
\deld_{\rm D}(\tau,\sigma)&= \tau-\theta(\tau-\sigma)\\
\ddeld_{\rm D}(\tau,\sigma)&= 1-\delta(\tau-\sigma)\\
\dddel_{\rm D}(\tau,\sigma)&=\delta(\tau-\sigma)\;.
\end{align}
Concerning the fermionic fields $\psi^{a}_{i}(\tau)$, we list here their propagator
\begin{equation} \label{AF}
\langle \psi^{a}_{i}(\sigma)\bar{\psi}^{b}_{j}(\tau)\rangle =\delta^{ab}\del_{ \rm{AF}}\,_{ij}(\tau,\sigma) =\delta^{ab}\left(
\begin{array}{cc}
 F(z,\tau,\sigma) & 0 \\
 0 & F(\omega,\tau,\sigma) \\
\end{array}
\right)
\end{equation}
where each entry in the matrix is a ${\cal N}=2$ fermionic propagator \cite{Bastianelli:2005vk} defined as
\begin{align}
F(z,\tau,\sigma) &=z^{-(\tau-\sigma)}\left(\frac{1}{z+1}\right)\left( z\, \theta(\tau-\sigma)-\theta(\sigma-\tau)\right)\\
F(\omega,\tau,\sigma) &=\omega^{-(\tau-\sigma)}\left(\frac{1}{\omega+1}\right)\left( \omega\, \theta(\tau-\sigma)-\theta(\sigma-\tau)\right)\ .
\end{align}
Next, we have the ghost variable propagators, defined as
\begin{align}
\langle a^{\mu}(\tau) a^{\nu}(\sigma)\rangle &= 2T g^{\mu\nu}(x_{0})\del_{_{\rm gh}}(\tau,\sigma)=2T g^{\mu\nu}(x_{0})\delta(\tau-\sigma)\\[1mm]
\langle b^{\mu}(\tau) c^{\nu}(\sigma)\rangle &=-4T g^{\mu\nu}(x_{0})\del_{_{\rm gh}}(\tau,\sigma) =-4T g^{\mu\nu}(x_{0})\delta(\tau-\sigma)\;.
\end{align}
As previously stated, the calculation involving these propagators may result in products/derivatives of delta distributions, which are ill-defined, but also divergent quantities such as $\delta(\tau,\tau)$, therefore one needs to regularize the path integral. In the present work, we chose worldline dimensional regularization (DR), which consists in continuing the compact time direction with the addition of $d$ non-compact extra dimensions, i.e. extending the space $\tau\in[0,1]\rightarrow t^\alpha=\left(\tau,\mathbf{t}\right)\in[0,1]\times\mathbb{R}^n$. More details can be found for example in \cite{Bastianelli:2006rx}; here we list the dimensional regularized expression of propagators, exploited during intermediate steps when performing computations. The $d+1$ extended propagators read
\begin{align}
\del_{\rm D}(t,s) &= \int \frac{d^{d}k}{(2\pi)^{d}}\sum_{m=1}^{\infty}\frac{-2}{(\pi m)^{2}+\mathbf{k}^{2}}\sin(\pi m \tau)\sin(\pi m \sigma)\,e^{i \mathbf{k}\cdot(\mathbf{t}-\mathbf{s})} \\
\del_{\rm gh}(t,s)&=\int \frac{d^{d}k}{(2\pi)^{d}}\sum_{m=1}^{\infty} 2\sin(\pi m\tau)\sin(\pi m \sigma) e^{i \mathbf{k}\cdot(\mathbf{t}-\mathbf{s})}=\delta(\tau-\sigma)\delta^{d}(t-s) \\
F(\theta,t,s)&=-i\int \frac{d^{d}k}{(2\pi)^{d}}\sum_{r\in Z+\frac{1}{2}}\frac{2\pi r\gamma^{0}+k\cdot \gamma-\theta}{(2\pi r)^{2}+\mathbf{k}^{2}-\theta^{2}}\, e^{2\pi i r(\tau-\sigma)}e^{i\mathbf{k}\cdot (\mathbf{t}-\mathbf{s})}
\end{align}
where in the dimensional regularized expressions one has $\mathbf{t}$ as $d$-dimensional vector and $\gamma^\alpha$ are the Dirac matrices in the $(d+1)$-dimensional extended space. Each extended worldline propagator satisfies a generalization of its own one-dimensional Green equation
\begin{align} \label{greenEQ}
\partial_{\mu}\partial^{\mu}\del_{\rm D}(t,s) &= \delta^{d+1}(t-s)\\
\left(\rpartial + i\theta\right)F(\theta,t,s)& = \delta_{\rm AF}(\tau-\sigma)\delta^{d}(t-s) \label{B16} \\
F(\theta,t,s) \left( -\lpartial +i\theta\right) &=\delta_{\rm AF}(\tau-\sigma)\delta^{d}(t-s)\ , \label{greenEQ2}
\end{align}
where $\delta_{\rm AF}$ is the delta distribution acting on antiperiodic functions on $[0,1]$ and where a slashed derivative is the usual contraction between derivative and gamma matrices. For computational purposes, the index contractions in $d+1$ dimensions serve mostly as a bookkeeping device to keep track of which derivative can be contracted to which vertex to produce the $(d+1)$-dimensional delta function. The delta functions in (\ref{greenEQ})--\eqref{greenEQ2} are only to be used in $d+1$ dimensions; then, by using partial integration one casts the various loop integrals in a form that can be computed by sending $d\to 0$ first. At this stage, one can use the one-dimensional propagators \eqref{DBC}--\eqref{AF}, and $\gamma^0=1$ (with no extra factors arising from the Dirac algebra in $d+1$ dimensions). 

%################################################################################################################
\subsection{Analisys of perturbative contributions}\label{appendixB2}
%################################################################################################################

In this appendix, we will give the essential details on the evaluation of the path integral average and its subsequent modular integration \eqref{2.16}, which produces the Seleey-DeWitt coefficients to be inserted in \eqref{series}. As anticipated in Section \ref{sec2.3}, in order to find all the possible contributions to order $T^3$ we have to expand the exponential with the interacting action \eqref{S-int}, written more compactly as in \eqref{10} which in particular carries the following vertices 
\begin{align}
&S_{\rm K1}=\int d\tau \; \frac{1}{3}R_{\alpha\mu\nu\beta}\, q^{\alpha}q^{\beta}\left( \dot{q}^{\mu}\dot{q}^{\nu} + \text{\rm gh}\right) \hskip.2cm ; \hskip.2cm S_{\rm C1} = \int d\tau \; \frac{1}{2}R_{\alpha\mu ab}\, \dot{q}^{\mu}q^{\alpha}\, \bar{\psi}^{a}\cdot \psi^{b}\\
&DS_{\rm C1}=\int d\tau \; \frac{1}{3}\nabla_{\alpha}R_{\beta\mu ab}\,\dot{q}^{\mu}q^{\alpha}q^{\beta} \, \bar{\psi}^{a}\cdot \psi^{b} \hskip.2cm ; \hskip.2cm D^{2}S_{\rm C1}= \int d\tau \; \frac{1}{8}\nabla_{\alpha}\nabla_{\beta}R_{\gamma\mu ab}\, \dot{q}^{\mu}q^{\alpha} q^{\beta}q^{\gamma}\, \bar{\psi}^{a}\cdot \psi^{b} \\
&S_{\rm C2} =\int d\tau \; \frac{1}{24}R^{\tau}{}_{\alpha\beta\mu}R_{\gamma \tau ab}\,
 \dot{q}^{\mu}q^{\alpha}q^{\beta}q^{\gamma}\, \bar{\psi}^{a}\cdot \psi^{b} \hskip.2cm ; \hskip.2cm S_{\rm F}= \int d\tau \; R_{abcd}\, \bar{\psi}^{a}\cdot \psi^{b}\bar{\psi}^{c}\cdot \psi^{d} \\
&DS_{\rm F}=\int d\tau \; q^\alpha \nabla_{\alpha}R_{abcd} \, \bar{\psi}^{a}\cdot \psi^{b}\bar{\psi}^{c}\cdot \psi^{d} \hskip.2cm ; \hskip.2cm D^{2}S_{\rm F}= \int d\tau \; \frac{1}{2} q^\alpha q^\beta \nabla_{\alpha}\nabla_{\beta}R_{abcd}\, \bar{\psi}^{a}\cdot \psi^{b}\bar{\psi}^{c}\cdot \psi^{d} \\
&S_{\rm V}=-\int d\tau \; \Omega R\ . \label{B.22}
\end{align}
Through Wick contractions they result in a plethora of different terms, most of which either give rise to disconnected diagrams (so that they are easily taken care of afterward, as connected diagrams exponentiate and there is no need to compute them anew) or can easily be shown to vanish. To start with, we see that the vertex $S_{\rm V}$ in \eqref{B.22} does not depend on $q$, $\psi$ nor on $\bar \psi$, thus it can be taken out of the path integral and it remains exponentiated (as will be all connected worldline diagrams). Let's see some other illustrative examples. For instance, a disconnected diagram arises from the term
\begin{equation}
-\frac{1}{8}T \langle S_{\rm K1}S_{\rm F}^2\rangle \sim \disc \sim \text{disconnected}\;.
\end{equation}
Other diagrams can be shown to be zero exploiting (anti)symmetry of the tensor structures and/or of the resulting propagators, like
\begin{align}
-\frac{1}{32 T^2} \langle S_{\rm K1}^2 S_{\rm C1}\rangle &\sim R_{\mu\nu ab}\langle\bar{\psi}^{a}\cdot \psi^{b} \rangle=0 \\
\frac{1}{4}\langle S_{\rm K1} S_{\rm C1}S_{\rm F}\rangle &\sim R_{ab}R_{\mu\nu}{}^{ab}R_{\alpha\rho\sigma\beta}=0\ .
\end{align}
Lastly, others are simply zero after explicitly calculating the integrals, as happens for 
$\langle S_{\rm C2}S_{\rm F}\rangle $.

We report now the surviving contributions \eqref{terms} in full glory:
\begin{align*}
\langle e^{-S_{\rm int}} \big \rangle \Big|_{T^3} 
&= \langle S_{\rm KIN} \rangle \tag{A}\\
&\phantom{=}-\frac{1}{96T}R_{\alpha\mu\nu\beta}R_{\gamma\lambda ab}R_{\delta\epsilon cd}\int_{012}
\langle \dot{q}^{\mu}_{0}\dot{q}^{\nu}_{0}q^{\alpha}_{0}q^{\beta}_{0}\dot{q}^{\lambda}_{1}q^{\gamma}_{1}\dot{q}^{\epsilon}_{2}q^{\delta}_{2}
\rangle \langle \bar{\psi}^{a}_{1}\cdot \psi_{1}^{b} \bar{\psi}^{c}_{2}\cdot \psi_{2}^{d} \rangle \tag{B} \\
&\phantom{=}+\frac{1}{48}R_{\alpha\mu ab}R^{\tau}{}_{\beta\lambda\nu}
R_{\rho \tau cd}\int_{01}\langle \dot{q}^{\mu}_{0}q^{\alpha}_{0}\dot{q}^{\nu}_{1}q^{\beta}_{1}q^{\lambda}_{1}q^{\rho}_{1}\rangle \langle \bar{\psi}^{a}_{0}\cdot \psi_{0}^{b} \bar{\psi}^{c}_{1}\cdot \psi_{1}^{d}\rangle \tag{C} \\
\tag{D} \label{D} &\begin{aligned} 
\begin{split}
\phantom{=}+\frac{1}{18}\nabla_{\alpha}R_{\beta\mu ab}\nabla_{\lambda}R_{\rho\nu cd}\int_{01} \langle \dot{q}^{\mu}_{0}q^{\alpha}_{0}q^{\beta}_{0}\dot{q}^{\nu}_{1}q^{\lambda}_{1}q^{\rho}_{1}\rangle \, \langle \bar{\psi}^{a}_{0}\cdot \psi^{b}_{0}\bar{\psi}^{c}_{1}\cdot \psi^{d}_{1}\rangle \\
\phantom{=} +\frac{1}{16}\nabla_{\alpha}\nabla_{\beta}R_{\lambda\mu ab} R_{\rho \nu cd}\int_{01} \langle \dot{q}^{\mu}_{0}q^{\alpha}_{0}q^{\beta}_{0}q^{\lambda}_{0}\dot{q}^{\nu}_{1}q^{\rho}_{1}\rangle \, \langle \bar{\psi}^{a}_{0}\cdot \psi^{b}_{0}\bar{\psi}^{c}_{1}\cdot \psi^{d}_{1}\rangle 
\end{split}
\end{aligned}\\
&\phantom{=} +\frac{T}{8}R_{\alpha\mu ab}R_{\beta\nu cd}R_{efgh}\int_{012}\langle \dot{q}^{\mu}_{0}\dot{q}^{\nu}_{1}q^{\alpha}_{0}q^{\beta}_{1}\rangle \langle \bar{\psi}^{a}_{0}\cdot \psi^{b}_{0}\bar{\psi}^{c}_{1}\cdot\psi^{d}_{1}\psi^{e}_{2}\cdot \bar{\psi}^{f}_{2}\psi^{g}_{2}\cdot \bar{\psi}^{h}_{2} \rangle \tag{E}\\
&\phantom{=}+\frac{1}{6}R_{abcd}R_{efgh}R_{lmno}\int_{012}\langle \bar{\psi}^{a}_{0}\cdot \psi^{b}_{0}\, \bar{\psi}^{c}_{0}\cdot \psi^{d}_{0}
\bar{\psi}^{e}_{1}\cdot \psi^{f}_{1}\, \bar{\psi}^{g}_{1}\cdot \psi^{h}_{1} \bar{\psi}^{l}_{2}\cdot \psi^{m}_{2}\, \bar{\psi}^{n}_{2}\cdot \psi^{o}_{2}\rangle \tag{F} \\
\tag{G} \label{G} &\begin{aligned} 
\begin{split}
&\phantom{=}+ \frac{T^2}{2} \nabla_{\alpha}R_{abcd}\nabla_{\beta}R_{efgh} \int_{01} \langle \dot{q}^{\alpha}_{0}\dot{q}^{\beta}_{1}\rangle \langle \bar{\psi}^{a}_{0}\cdot \psi^{b}_{0}\, \bar{\psi}^{c}_{0}\cdot \psi^{d}_{0}
\bar{\psi}^{e}_{1}\cdot \psi^{f}_{1}\, \bar{\psi}^{g}_{1}\cdot \psi^{h}_{1} \rangle \\
&\phantom{=} +\frac{T^2}{2}\nabla_{\alpha}\nabla_{\beta}R_{abcd}R_{efgh}\int_{01} \langle \dot{q}^{\alpha}_{0}\dot{q}^{\beta}_{0}\rangle \langle \bar{\psi}^{a}_{0}\cdot \psi^{b}_{0}\, \bar{\psi}^{c}_{0}\cdot \psi^{d}_{0}
\bar{\psi}^{e}_{1}\cdot \psi^{f}_{1}\, \bar{\psi}^{g}_{1}\cdot \psi^{h}_{1} \rangle\ .
\end{split}
\end{aligned}
\end{align*}
We decided to keep both the ghost fields and the term $S_{\rm KIN}$ implicit so as not to further burden the notation, and we combined the two terms of pure spin connection with covariant derivatives $\frac{1}{2}DS_{\rm C1}^{2}+D^{2}S_{\rm C1}S_{\rm C1}$ in eq. \eqref{D} and the two terms of the Taylor expansion of the four-fermions action $\frac{1}{2}T^2DS_{\rm F}^2 +T^2D^{2}S_{\rm F}S_{\rm F}$ in eq. \eqref{G}, for reasons that will become clear shortly thereafter. Below we illustrate the main steps of the calculation for each contribution. 
%-----------------------------------------------------------------------
\begin{enumerate}[label=(\Alph*)]
%-----------------------------------------------------------------------
% A
\item The first contribution we called $S_{\rm KIN}$ to indicate economically the sum of all terms arising \emph{only} from the pure kinetic part of the interacting action. In the notation previously presented it would read
\begin{equation}
S_{\rm KIN}=-\frac{1}{384 T^3}S_{\rm K1}^3+\frac{1}{32 T^2}DS_{\rm K1}^2+\frac{S_{\rm K1}}{16T^2}\left( S_{\rm K2}+D^2S_{\rm K1} \right) -\frac{1}{4T}S_{\rm K3}\ .
\label{B27}
\end{equation}
It has already been computed in \cite{Bastianelli:2000dw} (eq.\ 20) and can be read out, translated in our basis \eqref{E3}, from the exponential of connected diagrams as
\begin{equation}
\langle S_{\rm KIN} \rangle=\frac{T^{3}}{7!}	\left(	-\frac{16}{9}\frac{\mathcal{E}_{1}}{D^{2}}+\frac{2}{3}\frac{\mathcal{E}_{2}}{D}+\frac{17}{9}\mathcal{E}_{3}+\frac{28}{9}\mathcal{E}_{4}	\right)\ .
\end{equation}
%-----------------------------------------------------------------------
% B
\item Regarding the contribution obtained by coupling the bosonic kinetic term and the spin connection, once the contractions have been evaluated, one can use integration by parts (IBP) and DR to reduce integrals to a set of independent ones in $z$, modulo their $\omega$ partner.\footnote{This should also be understood for forthcoming worldline integrals when the fermionic integrands depend \emph{only} on the Wilson variable $z$.} We list here such integrals
\begin{align}
&\int_{012} \ddel(\tau,\sigma)\deld(\tau,\sigma)\ddel(\tau,\rho)\deld(\tau,\rho)F(z,\sigma,\rho)F(z,\rho,\sigma)=-\frac{z}{120\left(1+z	\right)^{2}} \\
&\int_{012} \ddeld(\tau,\rho)\ddel(\tau,\tau)\deld(\tau,\sigma)\del(\sigma,\rho)F(z,\sigma,\rho)F(z,\rho,\sigma)=-\frac{1}{720}\frac{z}{(z+1)^{2}}\ ,
\end{align}
so that the final result for $-\frac{1}{8T}\langle S_{\rm K1}S^{2}_{\rm SC1} \rangle$ reads
\begin{equation}
T^{3}\frac{\mathcal{E}_{2}}{D}\left[-\frac{1}{90}\frac{z}{(z+1)^{2}}-\frac{1}{90}\frac{\omega}{(\omega+1)^{2}}		\right]+T^{3}\mathcal{E}_{3}\left[-\frac{z}{60\left(1+z	\right)^{2}}-\frac{\omega}{60\left(1+\omega	\right)^{2}}\right]\ .
\end{equation}
%-----------------------------------------------------------------------
% C
\item The contribution of pure spin connection has already been evaluated in Section \ref{sec2.3}; we report here the result for completeness:
\begin{equation}
\langle S_{\rm C1}S_{\rm C2} \rangle=\left(\frac{1}{180}\frac{\mathcal{E}_{2}}{D}+\frac{1}{120}\mathcal{E}_{3}\right)\left[\frac{z}{(z+1)^{2}}+\frac{\omega}{(\omega+1)^{2}}\right]\ .
\end{equation}
%-----------------------------------------------------------------------
% D
\item The contribution of pure spin connection with covariant derivatives is made up of the sum of two pieces: using Einstein manifolds simplifications one can collect the whole result, with only the following diagrams to be evaluated
\begin{align}
&\int_{01} \ddeld(\sigma,\tau)\del(\sigma,\tau)\del(\sigma,\tau)F(z,\sigma,\tau)F(z,\tau,\sigma) = \frac{z}{45(z+1)^{2}}\\
&\int_{01} \ddeld(\sigma,\tau)\del(\sigma,\tau)\del(\sigma,\sigma)F(z,\sigma,\tau)F(z,\tau,\sigma)=\frac{z}{60(z+1)^{2}}\ ,
\end{align}
getting the final result
\begin{equation}
\left\langle \frac{1}{2}DS_{\rm C1}^{2}+D^{2}S_{\rm C1}S_{\rm C1} \right\rangle=-\frac{T^{3}}{360}\left[ \frac{z}{(z+1)^{2}}+\frac{\omega}{(\omega+1)^{2}}\right]\left(2\frac{\mathcal{E}_{2}}{D}-\mathcal{E}_{3}+4\mathcal{E}_{4}\right)\ .
\end{equation}
%-----------------------------------------------------------------------
% E
\item This is produced from the coupling of spin connection to the four-fermions vertex. The list of independent diagrams one needs to evaluate is
\begin{align}
&\int_{012} \ddeld(\sigma,\tau)\del(\sigma,\tau)F(z,\rho,\tau)F(z,\tau,\rho)F(z,\rho,\sigma)F(z,\sigma,\rho)=-\frac{z^{2}}{12(z+1)^{4}}\\
&\int_{012} \ddeld(\sigma,\tau)\del(\sigma,\tau)F(z,\rho,\tau)F(z,\tau,\sigma)F(z,\sigma,u)F(z,\rho,\rho) =-\frac{z(z-1)^{2}}{48(z+1)^{4}}\\
&\int_{012} \ddel(\sigma,\tau)\deld(\sigma,\tau)F(z,\sigma,\rho)F(z,\rho,\sigma)F(\omega,\tau,\rho)F(\omega,u,\tau)=-\frac{z\omega}{12(z+1)^{2}(\omega+1)^{2}}\ ,
\end{align}
such that the final result reads
\begin{align}
\left\langle \frac{1}{2}T^{2}S_{\rm C1}^{2}S_{\rm F} \right\rangle &= T^{3}\frac{\mathcal{E}_{2}}{D}\left[\frac{z\left(z-1	\right)^{2}}{12 \left(z+1\right)^{4}} +	\frac{\omega\left(\omega-1	\right)^{2}}{12 \left(\omega+1\right)^{4}}	\right]\\
&\phantom{=}+T^{3}\mathcal{E}_{3}\left[	\frac{z^{2}}{12\left(z+1\right)^{4}}+\frac{\omega^{2}}{12\left(\omega+1\right)^{4}}+\frac{z\omega}{3\left(	z+1\right)^{2}\left(\omega+1	\right)^{2}} \right] \ .
\end{align}
%-----------------------------------------------------------------------
% F
\item The one arising from the pure fermionic vertex is quite tricky to evaluate at once. To simplify the calculation, it is possible to exploit the subtle double copy structure underlying the $\mathcal{N}=4$ spinning particle, rewriting this term as a sum of contributions coming from the two copies of the $\mathcal{N}=2$ particles, i.e. for the two values of the internal index $i=1,2$. This allows us to rewrite the
above term as
\begin{align}
\frac{1}{6} S_{\rm F}^{3} &= \frac{1}{6} S_{1}^{3}(z) +\frac{1}{6} S_{1}^{3}(\omega) +\frac{1}{6} S_{\rm mix}^{3} + \frac{1}{2}S_{1}(z)S_{2}^{2}(\omega)+\frac{1}{2}S_{1}(z)^{2}S_{2}(\omega) \\ 
&\phantom{=}+\frac{1}{2}S_{1}^{2}(z)S_{\rm mix}+\frac{1}{2}S_{1}(z)S_{\rm mix}^{2}+\frac{1}{2}S_{2}^{2}(\omega)S_{\rm mix}+\frac{1}{2}S_{2}(\omega)S_{\rm mix}^{2}+S_{1}(z)S_{2}(\omega)S_{\rm mix}\ .
\end{align}
In the previous expression, we defined for simplicity the actions
\begin{align}
S_{1}(z) &= \int d\tau\, TR_{abcd}\,\bar{\psi}^{1a}\psi_{1}^{b}\bar{\psi}^{1c}\psi_{1}\,^{d} \\
S_{\rm mix} &= \int d\tau \left( TR_{abcd}\,\bar{\psi}^{a1}\psi^{b}_{1}\bar{\psi}^{2c}\psi^{d}_{2} + 1\leftrightarrow 2\right)\ ,
\end{align}
where we explicated the flavor index. Once performed the contractions with Mathematica, based on the \emph{xTensor} package \cite{garcia_2002}, we further simplify the result by reducing tensor structures using Bianchi identity as follows
\begin{align}
\label{RiemID3} &R_{\alpha\beta}{}^{\rho\sigma}R^{\alpha\mu\beta\nu}R_{\rho\mu\sigma\nu} =\frac{1}{4}\mathcal{E}_3\\[.5em]
\label{RiemID4} &R_{\mu\alpha\nu\beta}R^{\mu\rho\nu\sigma}R^{\alpha}{}_{\sigma}{}^{\beta}{}_{\rho} =-\frac{1}{4}\mathcal{E}_3+\mathcal{E}_4\ .
\end{align}
Once evaluated the fermionic diagrams our final answer for $\langle\frac{1}{6}T^{3}S_{\rm F}^{3} \rangle$ is
\begin{align}
\begin{split}
&\phantom{=+}\frac{\mathcal{E}_{1}}{D^{2}} \left[-\frac{(-1 + z)^2 z (1 + (-8 + z) z)}{6 (1 + z)^6}-\frac{(-1 + \omega)^2 \omega (1 + (-8 + \omega) \omega)}{6 (1 + \omega)^6}\right] \\[.3em]
&\phantom{=}+\frac{\mathcal{E}_{2}}{D}\left[	-\frac{(z-1)^2 z^2}{(z+1)^6}-\frac{(\omega-1)^2 \omega^2}{(\omega+1)^6}-\frac{2 (-1 + z)^2 z \omega }{(1 + z)^4 (1 + \omega )^2} -\frac{2 z (-1 + \omega )^2 \omega }{(1 + z)^2 (1 + \omega )^4}\right] \\[.3em]
&\phantom{=}+\mathcal{E}_{3}\left[\frac{z^4-2 z^3+z^2}{6 (z+1)^6}+ \frac{\omega ^4-2 \omega ^3+\omega ^2}{6 (\omega +1)^6} -\frac{2 z^2 \omega \ }{(1 + z)^4 (1 + \omega )^2}- \frac{2 z \omega ^2 }{(1 + \ z)^2 (1 + \omega )^4} \right] \\[.3em]
&\phantom{=}+\mathcal{E}_{4}\left[	 -\frac{4 \omega ^3}{3 (\omega +1)^6}-\frac{4 z^3}{3 (z+1)^6}-\frac{4 \omega z}{3(\omega +1)^2 (z+1)^2(\omega+1)^{2}} \right]\ .
\end{split}
\end{align}
%-----------------------------------------------------------------------
% G
\item Finally, one has the term coming from the Taylor expansion of the four-fermions interaction, made of two pieces. Just as in the \eqref{D} case, one can collect them into a single one working on the tensorial structure and using Einstein manifolds simplifications. The resulting diagrams to be evaluated are
\begin{align}
&\int_{01} \del(\tau ,\tau ) F(z,\sigma ,\tau )F(z,\sigma ,\tau ) F(z,\tau ,\sigma ) F(z,\tau ,\sigma )=-\frac{z^2}{6 (z+1)^4}\\
&\int_{01} \del (\tau ,\tau ) F(z,\sigma ,\tau ) F(z,\tau ,\sigma )F(\omega ,\sigma ,\tau ) F(\omega ,\tau ,\sigma )=-\frac{\omega z}{6 (\omega +1)^2 (z+1)^2}\ ,
\end{align}
since the partner diagrams $\del (\tau,\tau )\rightarrow \del (\tau,\sigma )$ can easily be seen to be proportional to the above. The final answer for $\langle \frac{1}{2}T^2DS_{\rm F}^2 +T^2D^{2}S_{\rm F}S_{\rm F} \rangle$ is then 
\begin{equation}
T^{3}\left[	\frac{z^{2}}{12\left(z+1\right)^{4}}+\frac{\omega^{2}}{12\left(\omega+1\right)^{4}}+\frac{z\omega}{3\left(	z+1\right)^{2}\left(\omega+1	\right)^{2}} \right] \left(2\frac{\mathcal{E}_{2}}{D}-\mathcal{E}_{3}+4\mathcal{E}_{4}\right)\ .
\end{equation}
%-------------------------------------------------------------------------------
\end{enumerate}
%-------------------------------------------------------------------------------

%################################################################################################################
\subsection{Summing up all the pieces}
%################################################################################################################

We can collect the result of all the computations above as follows
\begin{equation}
\alpha_{3}(z,\omega,D) = c_{0}(z,\omega)\,\frac{\mathcal{E}_{1}}{D^{2}}+c_{1}(z,\omega)\,\frac{\mathcal{E}_{2}}{D} +c_{2}(z,\omega)\, \mathcal{E}_{3}
+c_{3}(z,\omega)\,\mathcal{E}_{4}\ ,
\end{equation}
with the coefficients
\begin{align}
\begin{split}
c_{0}(z,\omega)&= -\frac{1}{2835}
-\frac{z \left(z^2-8 z+1\right) (z-1)^2}{6 (z+1)^6}-\frac{(\omega -1)^2 \omega \left(\omega ^2-8 \omega +1\right)}{6 (\omega +1)^6}
\end{split}\\[.5em]
\begin{split}
c_{1}(z,\omega)&=\frac{1}{7560}-\frac{z}{90 (z+1)^2}-\frac{\omega }{90 (\omega +1)^2}+\frac{z (z-1)^2}{12 (z+1)^4}+\frac{(\omega -1)^2 \omega }{12 (\omega +1)^4}+\frac{z^2}{6 (z+1)^4}+\frac{\omega ^2}{6 (\omega +1)^4} \\[.3em]
&\phantom{=}-\frac{z^2(z-1)^2}{(z+1)^6}-\frac{(\omega -1)^2 \omega ^2}{(\omega
 +1)^6}-\frac{2 \omega z (z-1)^2}{(\omega +1)^2 (z+1)^4}-\frac{2 (\omega -1)^2 \omega z}{(\omega
 +1)^4 (z+1)^2}+ \frac{2 z\omega}{3\left( z+1 \right)^2\left( \omega +1\right)^2} 
\end{split}\\[.5em]
\begin{split}
c_{2}(z,\omega)&=\frac{17}{45360} -\frac{z}{180 (z+1)^2}-\frac{\omega }{180 (\omega +1)^2} +\frac{(z-1)^2 z^2}{6(z+1)^6} +\frac{(\omega -1)^2 \omega ^2}{6 (\omega +1)^6} \\
&\phantom{=}-\frac{2 \omega z^2}{(\omega +1)^2 (z+1)^4}-\frac{2 \omega ^2 z}{(\omega +1)^4(z+1)^2}
\end{split}\\[.5em]
\begin{split}
c_{3}(z,\omega)&=\frac{1}{1620}-\frac{z}{90(z+1)^2}-\frac{\omega }{90 (\omega +1)^2}+ \frac{z^2}{3 \left( z+1 \right)^4} + \frac{\omega^2}{3 \left( \omega+1 \right)^4} + -\frac{4 \omega ^3}{3 (\omega +1)^6}-\frac{4 z^3}{3 (z+1)^6}\ .
\end{split}
\end{align}
At this point, in order to get the full $T^3$ correction, we first need to recall the connected diagrams coming from the first- and second-order expansion of the path integral, namely
\begin{align}
\alpha_{1}(z,\omega,D)&=\left(\frac{5}{12} +\Omega-\frac{z}{(z+1)^{2}}-\frac{\omega}{(\omega+1)^{2}}\right) R \\[.5em]
\alpha_{2}(z,\omega,D)&=\left(
-\frac{1}{180}+\frac{1}{2}\left[\frac{z(z-1)^{2}}{(z+1)^{4}}+\frac{\omega(\omega-1)^{2}}{(\omega+1)^{4}}
\right]\right)\frac{R^{2}}{D} \nonumber\\[.3em]
&\phantom{=}+\left(\frac{1}{180}+\frac{1}{2} \left[\frac{\omega ^2}{(\omega +1)^4}+\frac{z^2}{(z+1)^4}+\frac{4 \omega z}{(\omega +1)^2 (z+1)^2}\right]-\frac{1}{12} \left[\frac{\omega }{(\omega
 +1)^2}+\frac{z}{(z+1)^2}\right] \right)R_{\mu\nu\rho\sigma}^{2}\nonumber \\[.3em]
&\equiv \beta_{1}(z,\omega)\, \frac{R^{2}}{D}+\beta_{2}(z,\omega)\,R_{\mu\nu\rho\sigma}^{2}\ ,
\end{align}
which are in complete accordance with the results of \cite{Bastianelli:2019xhi, Bastianelli:2022pqq} and indeed reproduce the $a_1$ and $a_2$ coefficients \eqref{a1}--\eqref{a2}. Now we can move to the exponentiation of all the connected diagrams, including the new results. The expectation value $\big\langle e^{-S_{\rm int}}\big\rangle$ in \eqref{2.16} can be compactly written as
\begin{align}
\begin{split}
\Big\langle e^{-S_{\rm int}}\Big\rangle &=\exp\Bigg[
T\,\alpha_{1}\, R+T^{2}\,\left(\beta_{1}\, \frac{R^{2}}{D}+\beta_{2}\,R_{\mu\nu\rho\sigma}^{2}	\right) \\
&\phantom{=}+ T^{3}\,\left( c_{0}\,\frac{R^{3}}{D^{2}}+c_{1}\,\frac{R\, R_{\mu\nu\rho\sigma}^{2}}{D}+ c_{2}\,R_{\mu\nu\rho\sigma}R^{\rho\sigma\alpha\beta}R_{\alpha\beta}{}^{\mu\nu}
+ c_{3}\,R_{\alpha\mu\nu\beta}R^{\mu\rho\sigma\nu}R_{\rho}{}^{\alpha\beta}{}_{\sigma} \right) + {\cal O}(T^4) \Bigg]\ .
\end{split}
\end{align}
The final step consists of Taylor expanding the exponential to the desired order, namely
\begin{equation}
\Big\langle e^{-S_{\rm int}}\Big\rangle\Big|_{T^{3}}= T^{3}\left[ \left( \frac{\alpha_{1}^{3}}{6}+\frac{\alpha_{1}\beta_{1}}{D}+\frac{c_{0}}{D^{2}}\right)\,\mathcal{E}_1+\left( \alpha_{1}\beta_{2}+\frac{c_{1} }{D}\right)\,\mathcal{E}_2 +c_{2}(z,\omega)\, \mathcal{E}_{3} +c_{3}(z,\omega)\,\mathcal{E}_{4} \right]+ \mathcal{O}(T^{4})\ .
\end{equation}
We obtained our final answer for the path integral average \eqref{2.16}. In the end, one is left only with performing the modular integration, i.e. the double expectation value $\langle \hskip -.05 cm \langle e^{-S_{\rm int}} \rangle \hskip -.05cm \rangle$ defined in \eqref{double}. In doing so, we can choose the appropriate measure $P(z,\omega)$ to project only on the ghost, graviton, or total coefficients. This gives our final results of Section \ref{sec2.4}, in particular the newly-computed coefficient
\begin{align}
a_{3}(D)&= \frac{35 D^4-147 D^3-3670 D^2-13560D-30240}{90720 D^2}\;\mathcal{E}_1 +\frac{7 D^3-230 D^2+3357 D+12600}{15120 D}\;\mathcal{E}_2+ \nonumber\\[.3em] 
&\phantom{=}+ \frac{17 D^2-555 D-15120}{90720}\;\mathcal{E}_3 +\frac{D^2-39 D-1080}{3240}\; \mathcal{E}_4\ .
\end{align}

%################################################################################################################
%################################################################################################################

%################################################################################################################
%################################################################################################################
\section{Heat Kernel computations}\label{appendixC}

The fourth heat kernel coefficient for the ghost field is computed from the general formula \eqref{HK3}, by performing the substitutions \eqref{GhostSost}. It is convenient to write
\begin{equation}\label{GhostHK-split}
\alpha_3^{\rm gh}(x) \equiv \frac{1}{7!}\bm{A}_{\rm gh}[R, R_{\mu\nu}, R_{\mu\nu\rho\sigma}] + \frac{2}{6!}\bm{B}_{\rm gh}[R, R_{\mu\nu}, R_{\mu\nu\rho\sigma}, \Omega_{\mu\nu}, V]\ ,
\end{equation}
where $\bm{A}_{\rm gh}$ and $\bm{B}_{\rm gh}$ are two (involved) functions of the metric invariants and of the gauge field strength, as reported in \eqref{HK3}. Let us start from $\bm{A}_{\rm gh}$: six of the first eight terms vanish identically since they are all proportional to covariant derivatives of $R$, $R_{\mu\nu}$ or $R_{\mu\nu\rho\sigma}$, except from the two proportional to $(\nabla_\alpha R_{\mu\nu\rho\sigma})^2$ and $R_{\mu\nu\rho\sigma}\nabla^2R^{\mu\nu\rho\sigma}$. Recalling that $\Tr{[\delta^\mu_\nu]}=D$ and using \eqref{EinsteinCond}, we are therefore left with
\begin{align}
\begin{split}\label{Agh}
\Tr\left[\bm{A}_{\rm gh}\right] &= D\left(3R_{\mu\nu\rho\sigma}\nabla^2R^{\mu\nu\rho\sigma}-\frac{208}{9}R_{\mu}{}^{\nu}R_{\nu}{}^{\sigma}R_{\sigma}{}^{\mu} + \frac{64}{3}R_{\mu\nu}R_{\rho\sigma}R^{\mu\rho\nu\sigma}-\frac{16}{3}R_{\mu\nu}R^\mu{}_{\rho\sigma\tau}R^{\nu\rho\sigma\tau}\right.\\[.3em]
&\phantom{=}\left.+\frac{44}{9}R_{\mu\nu}{}^{\rho\sigma}R_{\rho\sigma}{}^{\alpha\beta}R_{\alpha\beta}{}^{\mu\nu}+\frac{80}{9}R_{\mu\nu\rho\sigma}R^{\mu\alpha\rho\beta}R^{\nu}{}_{\alpha}{}^{\sigma}{}_{\beta}\right)\\[.3em]
&= -\frac{16}{9D}\mathcal{E}_1+\frac{2}{3}\mathcal{E}_2 +\frac{17D}{9}\mathcal{E}_3 +\frac{28D}{9}\mathcal{E}_4\ .
\end{split}
\end{align}
We now compute $\bm{B}_{\rm gh}$. Since $(\Omega_{\mu\nu})^\rho{}_{\sigma} = R_{\mu\nu}{}^{\rho}{}_{\sigma}$, the second term of $\bm{B}_{\rm gh}$ in \eqref{HK3} vanishes identically; moreover, as $V\propto R$, the same occurs for the last four, leaving us with
\begin{align}
\begin{split}\label{Bgh}
\Tr\left[\bm{B}_{\rm gh}\right] &= 4\Tr{\left[\Omega_{\mu\nu}\nabla^2\Omega^{\mu\nu}\right]}-12\Tr{\left[\Omega_{\mu}{}^{\nu}\Omega_{\nu}{}^{\sigma}\Omega_{\sigma}{}^{\mu} \right]}+ 6\Tr{\left[R_{\mu\nu\rho\sigma}\Omega^{\mu\nu}\Omega^{\rho\sigma}\right]}-4 \Tr{\left[R_{\mu\nu}\Omega^{\mu\sigma}\Omega^{\nu}{}_{\sigma}\right]}\\[.3em]
&= -\frac{4}{D}\mathcal{E}_2- 2\mathcal{E}_3 -4\mathcal{E}_4\ ,
\end{split}
\end{align}
where all the traces are computed by explicit substitution, according to \eqref{GhostSost},
\begin{align}
\Tr{\left[\Omega_{\mu}{}^{\nu}\Omega_{\nu}{}^{\sigma}\Omega_{\sigma}{}^{\mu}\right]} &= \Tr{\left[R^{\alpha\ \ \nu}_{\ \beta\mu}R^{\beta\ \ \sigma}_{\ \gamma\nu}R^{\gamma\ \ \mu}_{\ \delta\sigma}\right]} = R^{\alpha\ \ \nu}_{\ \beta\mu}R^{\beta\ \ \sigma}_{\ \gamma\nu}R^{\gamma\ \ \mu}_{\ \alpha\sigma} = -\mathcal{E}_4\\[.5em]
\Tr{\left[R_{\mu\nu\rho\sigma}\Omega^{\mu\nu}\Omega^{\rho\sigma}\right]} &= R_{\mu\nu\rho\sigma} \Tr{\left[R^{\alpha\ \mu\nu}_{\ \beta} R^{\beta\ \rho\sigma}_{\ \gamma}\right]} = R_{\mu\nu\rho\sigma}R^{\alpha\ \mu\nu}_{\ \beta} R^{\beta\ \rho\sigma}_{\ \alpha}= -\mathcal{E}_3\\[.5em]
\Tr{\left[R_{\mu\nu}\Omega^{\mu\sigma}\Omega^{\nu}{}_{\sigma}\right]} &= R_{\mu\nu} \Tr{\left[R^{\alpha\ \mu\sigma}_{\ \beta}R^{\beta\ \nu}_{\ \gamma\ \sigma}\right]} = R_{\mu\nu} R^{\alpha\ \mu\sigma}_{\ \beta}R^{\beta\ \nu}_{\ \alpha\ \sigma}= -\frac{1}{D}\mathcal{E}_2\ .
\end{align}
Going back to \eqref{GhostHK-split} we conclude that
\begin{equation}\label{GhostAlpha}
\Tr{\left[\alpha_3^{\rm gh}(x)\right]} = -\frac{1}{2835 D}\mathcal{E}_1+\frac{D-84}{7560 D}\mathcal{E}_2 +\frac{17 D-252}{45360}\mathcal{E}_3 +\frac{D-18}{1620}\mathcal{E}_4\ ,
\end{equation}
which however does not provide the full coefficient for the ghost, since we still have to add the term $\beta_3$ defined in \eqref{BetaValues}, which turns out to be
\begin{align}
\begin{split}
\beta_3^{\rm gh} &= \frac{1}{6}\left(\alpha_1^{\rm gh}\right)^3+\alpha_1^{gh}\alpha_2^{\rm gh} \\
&= \frac{1}{6}\left[\delta^\tau_\alpha\left(\frac{1}{6}+\frac{1}{D}\right)R\right]^3 + \left[\delta^\tau_\alpha\left(\frac{1}{6}+\frac{1}{D}\right)R\right]\left[\frac{1}{180}\left(R_{\mu\nu\rho\sigma}^2-R_{\mu\nu}^2\right)\delta^\alpha_\gamma+\frac{1}{12}\Omega_{\mu\nu}^2\right]\ ,
\end{split}
\end{align}
and taking the trace
\begin{equation}\label{Beta3Ghost}
\Tr{\left[\beta_3^{\rm gh}(x)\right]} = \frac{(5D^2+54D+180)(D+6)}{6480D^2}\mathcal{E}_1+\frac{(D+6)(D-15)}{1080D}\mathcal{E}_2\ .
\end{equation}
The overall result for the ghost is
\begin{align}
\begin{split}
\Tr{\left[a_3^{\rm gh}(x)\right]} &= \frac{35D^3+588D^2+3512D+7560}{45360D^2}\ \mathcal{E}_1 \\[.5em]
&\phantom{=} + \frac{7D^2-62D-714}{7560D}\mathcal{E}_2 +\frac{17D-252}{45360}\mathcal{E}_3 +\frac{D-18}{1620}\mathcal{E}_4
\end{split}
\end{align}
and corresponds to \eqref{a3gh}. To compute the fourth heat kernel coefficient for the graviton, we start again from the general formula \eqref{HK3} and perform the substitutions \eqref{GravitonSost}. It is once again more time convenient to split
\begin{equation}\label{GravitonHK-split}
\alpha_3^{\rm gr}(x) \equiv \frac{1}{7!}\bm{A}_{\rm gr}[R, R_{\mu\nu}, R_{\mu\nu\rho\sigma}] + \frac{2}{6!}\bm{B}_{\rm gr}[R, R_{\mu\nu}, R_{\mu\nu\rho\sigma}, \Omega_{\mu\nu}, V]
\end{equation}
as we did in \eqref{GhostHK-split} for the ghost. By inspecting more closely the substitution rules \eqref{GhostSost} and \eqref{GravitonSost}, though, it is clear that $\bm{A}_{\rm gr}$ and $\bm{A}_{\rm gh}$ differ only by the trace of the identity operator $\mathbbm{1}$. Using \eqref{Agh}, we find in particular
\begin{equation}\label{Agr}
\Tr\left[\bm{A}_{\rm gr}\right] = \frac{D+1}{2}\, \Tr\left[\bm{A}_{\rm gh}\right]
= (D+1)\left(-\frac{8}{9D}\mathcal{E}_1+ \frac{1}{3}\mathcal{E}_2 +\frac{17D}{18}\mathcal{E}_3 +\frac{14D}{9}\mathcal{E}_4\right)\ ,
\end{equation}
while for the computation of $\bm{B}_{\rm gr}$ we can repeat the previous observations, and consider the only non-vanishing terms
\begin{align}
\begin{split}\label{Bgr}
\Tr\left[\bm{B}_{\rm gr}\right] &= 4\Tr{\left[\Omega_{\mu\nu}\nabla^2\Omega^{\mu\nu}\right]}-12\Tr{\left[\Omega_{\mu}{}^{\nu}\Omega_{\nu}{}^{\sigma}\Omega_{\sigma}{}^{\mu}\right]}\\[.3em]
& \phantom{=}+ 6\Tr{\left[R_{\mu\nu\rho\sigma}\Omega^{\mu\nu}\Omega^{\rho\sigma}\right]}-4 \Tr{\left[R_{\mu\nu}\Omega^{\mu\sigma}\Omega^{\nu}{}_{\sigma}\right]}+30\Tr{\left[(\nabla_\mu V)^2\right]}\\[.5em]
&=-\frac{4(D+47)}{D}\ \mathcal{E}_2- 2(D-43)\ \mathcal{E}_3 -4(D+92)\ \mathcal{E}_4
\end{split}
\end{align}
where we need to compute the traces
\begin{align}
\Tr{\left[\Omega_{\mu}{}^{\nu}\Omega_{\nu}{}^{\sigma}\Omega_{\sigma}{}^{\mu}\right]} &= -\left(D+2\right) \mathcal{E}_4\\[.3em]
\Tr{\left[R_{\mu\nu\rho\sigma}\Omega^{\mu\nu}\Omega^{\rho\sigma}\right]} &= -(D+2)\ \mathcal{E}_3\\[.3em]
\Tr{\left[R_{\mu\nu}\Omega^{\mu\sigma}\Omega^{\nu}{}_{\sigma}\right]} &= -\frac{D+2}{D}\ \mathcal{E}_2\\[.3em]
\Tr{\left[\Omega_{\mu\nu}\nabla^2\Omega^{\mu\nu}\right]} &= -(D+2)\left(\frac{2}{D}\mathcal{E}_2-\mathcal{E}_3+4\mathcal{E}_4\right)\\[.3em]
\Tr{\left[(\nabla_\alpha \mathcal{V}_{\mu\nu}{}^{\rho\sigma})^2\right]} &= -3\left(\frac{2}{D}\mathcal{E}_2-\mathcal{E}_3+4\mathcal{E}_4\right)\ .
\end{align}
Going back to \eqref{GravitonHK-split} we conclude that
\begin{align}
\begin{split}\label{GravitonAlpha} \Tr{\left[\alpha_3^{\rm gr}(x)\right]} &=-\frac{D+1}{5670 D}\ \mathcal{E}_1+\frac{D^2-167 D-7896}{15120 D}\ \mathcal{E}_2\\[.5em]
&\hspace{.5cm} +\frac{17 D^2-487 D+21672}{90720}\ \mathcal{E}_3 +\frac{D^2-35 D-3312}{3240}\ \mathcal{E}_4\ .
\end{split}
\end{align}
The last step is to compute the term $\beta_3$ defined in \eqref{BetaValues}, which is the sum of the two terms:
\begin{align}
\frac{1}{6}\left(\alpha_1^{\rm gr}\right)^3 &= \frac{1}{6} \left(\frac{1}{216}R^3\delta_{\mu\nu}^{\ \ \ \alpha\beta} + \frac{1}{12}R^2\mathcal{V}_{\mu\nu}^{\ \ \ \alpha\beta} + \frac{1}{2}R \, \mathcal{V}_{\mu\nu}^{\ \ \ \rho\sigma}\mathcal{V}_{\rho\sigma}^{\ \ \ \alpha\beta}+\mathcal{V}_{\mu\nu}^{\ \ \ \rho\sigma}\mathcal{V}_{\rho\sigma}^{\ \ \ \lambda\tau}\mathcal{V}_{\lambda\tau}^{\ \ \ \alpha\beta}\right)\\
\alpha_1^{\rm gr}\alpha_2^{\rm gr} &= \left(\frac{1}{6}R\delta_{\mu\nu}^{\ \ \ \alpha\beta}+\mathcal{V}_{\mu\nu}^{\ \ \ \alpha\beta}\right) \left[\frac{1}{180}\left(R_{\mu\nu\rho\sigma}^2-\frac{1}{D}R^2\right)\delta_{\alpha\beta}^{\ \ \ \lambda\tau} + \frac{1}{12}\left(\Omega_{\rho\sigma}^2\right)_{\alpha\beta}^{\ \ \ \lambda\tau}\right]+\frac{1}{6}\mathcal{V}_{\mu\nu}^{\ \ \ \alpha\beta}\nabla^2\mathcal{V}_{\alpha\beta}^{\ \ \ \lambda\tau}\nonumber
\end{align}
where the following traces are to be computed:
\begin{align}
\Tr{\left[\delta_{\mu\nu}^{\ \ \ \alpha\beta}\right]} &=\frac{1}{2}D(D+1)\\[.3em]
\Tr{\left[\mathcal{V}_{\mu\nu}^{\ \ \ \alpha\beta}\right]} &= -R\\[.3em]
\Tr{\left[\mathcal{V}_{\mu\nu}^{\ \ \ \rho\sigma}\mathcal{V}_{\rho\sigma}^{\ \ \ \alpha\beta}\right]} &=3R^2_{\mu\nu\rho\sigma}\\[.3em]
\Tr{\left[\mathcal{V}_{\mu\nu}^{\ \ \ \rho\sigma}\mathcal{V}_{\rho\sigma}^{\ \ \ \lambda\tau}\mathcal{V}_{\lambda\tau}^{\ \ \ \alpha\beta}\right]} &= -8\mathcal{E}_4 - \mathcal{E}_3\\[.3em]
\Tr{\left[\mathcal{V}_{\mu\nu}^{\ \ \ \alpha\beta}\left(\Omega_{\rho\sigma}^2\right)_{\alpha\beta}^{\ \ \ \lambda\tau}\right]} &= \frac{2}{D}\mathcal{E}_2 +3\mathcal{E}_3\\[.3em]
\Tr{\left[\mathcal{V}_{\mu\nu}^{\ \ \ \alpha\beta}\nabla^2\mathcal{V}_{\alpha\beta}^{\ \ \ \lambda\tau}\right]} &= 3\left(\frac{2}{D}\mathcal{E}_2-\mathcal{E}_3+4\mathcal{E}_4\right)
\end{align}
leading to
\begin{align}
\begin{split}\label{Beta3Graviton}
\Tr{\left[\beta_3^{\rm gr}(x)\right]} &= \frac{5 D^3-D^2-186 D+72}{12960 D}\ \mathcal{E}^3_1\\[.5em]
&\phantom{=} + \frac{D^3-29 D^2+468 D+2520}{2160 D}\ \mathcal{E}_2^3 - \frac{5}{12}\mathcal{E}_3^3+\frac{2}{3}\mathcal{E}_4^3\ .
\end{split}
\end{align}
In the end,
\begin{align}
\begin{split}
\Tr{\left[a_3^{\rm gr}(x)\right]} &= \frac{35 D^3-7 D^2-1318 D+488}{90720 D}\ \mathcal{E}_1 +\frac{7 D^3-202 D^2+3109 D+9744}{15120 D}\ \mathcal{E}_2 \\[.5em]
&+ \frac{17 D^2-487 D-16128}{90720}\ \mathcal{E}_3 +\frac{D^2-35 D-1152}{3240}\ \mathcal{E}_4\ ,
\end{split}
\end{align}
which corresponds to \eqref{a3gr}. By summing the ghost and graviton coefficients according to \eqref{TotHKgen}, we end up with
\begin{align}
\Tr{\left[a_3(x)\right]} &= \frac{35 D^4-147 D^3-3670 D^2-13560 D-30240}{90720 D^2}\ \mathcal{E}_1 +\frac{7 D^3-230 D^2+3357 D+12600}{15120 D}\ \mathcal{E}_2 \nonumber\\[.5em]
&\phantom{=} +\frac{17 D^2-555 D-15120}{90720}\ \mathcal{E}_3+\frac{D^2-39 D-1080}{3240}\ \mathcal{E}_4\ ,
\end{align}
which corresponds to \eqref{a3}.

%################################################################################################################
%################################################################################################################

%################################################################################################################
%################################################################################################################
\section{Topological terms in even dimensions}\label{appendixD}

In order to display the cancellation of the one-loop divergences of pure gravity in four spacetime dimensions, and analogously their resilience in six, we exploit the Gauss-Bonnet theorem, which allows us to compute the Euler character $\chi_{\rm E}(\mathcal{M})$ of a manifold $\mathcal{M}$ as a volume integral of the 2-form $\mathcal{R}^{\mu\nu}\equiv R^{\mu\nu}_{\ \ \ \alpha\beta}\,\diff{x}^\alpha\wedge\diff{x}^\beta$, namely
\begin{equation}\label{EulerChar}
\chi_{\rm E}(\mathcal{M})=\frac{1}{(4\pi)^d}\int_{\mathcal{M}} \varepsilon_{\mu_1\nu_1\dots\mu_d\nu_d} \mathcal{R}^{\mu_1\nu_1}\wedge\dots\wedge\mathcal{R}^{\mu_d\nu_d}\ ,
\end{equation}
where $D\equiv 2d$ is the dimension of the manifold, assumed here to be even.\footnote{If $D$ is odd, the integral \eqref{EulerChar} vanishes, so that the theorem does not provide a useful way of computing $\chi_{\rm E}(\mathcal{M})$.} In local coordinates, \eqref{EulerChar} becomes
\begin{equation}\label{EulerChar2}
\chi_{\rm E}(\mathcal{M})=\frac{1}{2(4\pi)^d}\int \diff{^Dx}\sqrt{g}\ \frac{D!}{D} \delta^{\alpha_1}_{[\mu_1} \delta^{\beta_1}_{\nu_1}\dots\delta^{\alpha_d}_{\mu_d} \delta^{\beta_d}_{\nu_d]} R^{\mu_1\nu_1}_{\ \ \ \ \ \alpha_1\beta_1}\dots R^{\mu_d\nu_d}_{\ \ \ \ \ \alpha_d\beta_d}\ .
\end{equation}
It is possible to prove that $\chi_{\rm E}(\mathcal{M})$ defined in this way does not depend on the metric settled upon $\mathcal{M}$, and is fixed only by the global topology of the manifold. For $D=2$ ($d=1$), \eqref{EulerChar2} becomes
\begin{equation}\label{EulerD2}
\left. \chi_{\rm E}(\mathcal{M})\right|_{D=2}= \frac{1}{16\pi} \int \diff{^2x} \sqrt{g}\ R\ ,
\end{equation}
which is proportional to Einstein-Hilbert action, while at $D=4$ ($d=2$) the Euler character reads
\begin{equation}\label{EulerD4}
\left. \chi_{\rm E}(\mathcal{M})\right|_{D=4} = \frac{1}{32\pi^2} \int \diff{^4x} \sqrt{g}\ \left(R^2-4R_{\mu\nu}R^{\mu\nu}+R_{\mu\nu\rho\sigma}R^{\mu\nu\rho\sigma}\right)\ .
\end{equation}
On Einstein spaces the first two terms in the integrand of \eqref{EulerD4} cancel off, and therefore we are left with
\begin{equation}\label{EulerD4bis}
\left. \chi_E(\mathcal{M})\right|_{D=4} = \frac{1}{32\pi^2} \int \diff{^4x} \sqrt{g}\ R_{\mu\nu\rho\sigma}R^{\mu\nu\rho\sigma}\equiv \int \diff{^4x} \sqrt{g}\ E_4\ .
\end{equation}
Therefore, the third heat kernel coefficient $\Tr{\left[a_2(x)\right]}\propto R_{\mu\nu\rho\sigma}^2$ is proportional to the Euler density $E_4$ and hence is a total derivative, which can be neglected in the effective action. This result is no more true when $\Lambda\neq 0$, even if we drop the total derivative term corresponding to Euler density. 

In dimension $D=6$ ($d=3$) the Euler character is \cite{Groh2012}
\begin{equation}\label{EulerD6}
\left. \chi_{\rm E}(\mathcal{M})\right|_{D=6} = \frac{1}{384\pi^3} \int \diff{^6x} \sqrt{g}\ \left(4\mathcal{K}_1-48\mathcal{K}_2+64\mathcal{K}_4+96\mathcal{K}_5+12\mathcal{K}_3-96\mathcal{K}_6+16\mathcal{K}_7-32\mathcal{K}_8\right)\ ,
\end{equation}
which on Einstein spaces reduces to
\begin{equation}\label{EulerD6bis}
\left. \chi_{\rm E}(\mathcal{M})\right|_{D=6} = \frac{1}{384\pi^3} \int \diff{^6x} \sqrt{g}\ \left(\frac{4}{9}\mathcal{E}_1-4\mathcal{E}_2+16\mathcal{E}_3+32\mathcal{E}_4\right)\ .
\end{equation}
The condition $\Lambda=0$, that is $R=0$, forces $\mathcal{E}_1=\mathcal{E}_2=0$, so that \eqref{EulerD6bis} eventually becomes
\begin{equation}\label{EulerD6tris2}
\left. \chi_{\rm E}(\mathcal{M})\right|_{D=6} = \frac{1}{384\pi^3} \int \diff{^6x} \sqrt{g}\ \left(16\mathcal{E}_3+32\mathcal{E}_4\right)\equiv \int \diff{^6x} \sqrt{g}\ E_6\ .
\end{equation}
This shows that, up to a total derivative term
\begin{equation}
2 \mathcal{E}_4 =-\mathcal{E}_3\ . 
\end{equation}
Thus, at dimension $D=6$, even with $\Lambda=0$, the perturbative quantum gravity effective action is not free of logarithmic divergences. 

%################################################################################################################
%################################################################################################################

%################################################################################################################
%################################################################################################################
\addcontentsline{toc}{section}{References}
\printbibliography

%################################################################################################################
%################################################################################################################

\end{document}